\documentclass[prd,aps,a4paper,preprintnumbers,superscriptaddress,twocolumn,nofootinbib]{revtex4}
\usepackage{graphicx}
\usepackage{color}
\usepackage{dcolumn}
\usepackage{bm}
\usepackage{slashed}
\usepackage{amsmath}
\usepackage{latexsym}
\usepackage{amssymb}
\usepackage{mathrsfs}
\usepackage{amsfonts}
\usepackage{longtable}
\usepackage{physics}
\usepackage{url}
\usepackage{float}
\usepackage{hyperref}

\allowdisplaybreaks
\begin{document}

\preprint{IFT-UAM/CSIC-26-6}

\newcommand{\XL}[1]{\textcolor{blue}{[{\bf XL}: #1]}}% Xiaolin's comment
\def\Omgw{\Omega_{\mathrm{gw}}}
\def\Omseg{\Omega_{\mathrm{seg}}}
\def\Omref{\Omega_{\mathrm{ref}}}
\def\ee{\mathrm{e}}
\def\hh{\mathrm{h}}
\def\AA{\mathrm{A}}
\def\FF{\mathrm{F}}
\def\NN{\mathrm{N}}
\def\bb{\mathrm{b}}
\def\cc{\mathrm{c}}
\def\ss{\mathrm{s}}
\def\Gaus{\mathcal{G}}
\def\ii{\mathtt{i}}
\def\PD{\mathrm{PD}}
\def\Erfc{\mathrm{Erfc}}
\def\PFA{\mathrm{PFA}}
\def\Like{\mathcal{L}}
\def\BF{\mathcal{B}}
\def\Cov{\mathbb{C}}
\def\tot{\mathrm{tot}}
\def\Nseg{N_\mathrm{seg}}
\def\Mch{\mathcal{M}}
\def\seg{\mathrm{seg}}
\newcommand{\avg}[1]{\left\langle{#1}\right\rangle}
\newcommand{\phat}[1]{\hat{\pmb{#1}}}
\def\Tseg{T_{\mathrm{seg}}}
\def\Amp{\mathcal{A}}
\def\AmpC{\hat{\mathcal{A}}}
\def\PsiC{\hat{\Psi}}
\def\PN{\mathrm{PN}}
\def\Hz{\mathrm{Hz}}
\def\fref{f_{\mathrm{ref}}}
\def\fmin{f_{\min}}
\def\fmax{f_{\max}}
\def\EE{\mathrm{E}}
\def\Tseg{T_{\mathrm{seg}}}
\def\Var{\mathrm{Var}}
\def\eff{\mathrm{eff}}
\def\mix{\mathrm{mix}}
\def\hf{\tilde{h}}
\def\Hf{\tilde{H}}
\def\nf{\tilde{n}}
\def\sf{\tilde{s}}
\def\df{\tilde{d}}
\def\tlbd{\tilde{\lambda}}
\def\tgm{\tilde{\gamma}}
\def\Hubble{\mathcal{H}}
\def\hatOmg{\hat{\Omega}}
\def\Ngw{N_{\mathrm{gw}}}
\def\pOmg{\pmb{\Omega}}
\def\Gauss{\pmb{\mathrm{G}}}
\def\gw{\mathrm{gw}}
\def\Hpref{H_{\mathrm{M}}}
\def\lnmu{\ln{\mu}}
\def\sigln{\sigma_{\ln}}
\def\lnG{\ln{\mathcal{G}}}
\def\lnSG{\ln{\mathcal{G}_S}}
\def\skew{\mathrm{Skew}}
\def\kurt{\mathrm{Kurt}}
\def\erfc{\mathrm{Erfc}}
\def\sigdet{\sigma_{\mathrm{det}}}
\def\tGm{\tilde{\pmb{\Gamma}}}
\def\pgm{\pmb{\gamma}}
\def\HH{\mathbf{H}}

\title{Analyzing intermittent stochastic gravitational wave background I:\\
Effect of detector response}

\author{Xiaolin Liu}\email[Xiaolin Liu: ]{shallyn.liu@foxmail.com}
\affiliation{Instituto de Física Téorica UAM-CSIC, Universidad Autónoma de Madrid, Cantoblanco 28049 Madrid, Spain}
\author{Sachiko Kuroyanagi\footnote{corresponding author}} \email[Sachiko Kuroyanagi: ]{sachiko.kuroyanagi@csic.es}
\affiliation{Instituto de Física Téorica UAM-CSIC, Universidad Autónoma de Madrid, Cantoblanco 28049 Madrid, Spain}
\affiliation{Department of Physics, Nagoya University, Chikusa, Nagoya 464-8602, Japan}

\begin{abstract}
With the growing number of gravitational-wave detections, particularly from binary black hole mergers, there is increasing anticipation that an astrophysical background, formed by an ensemble of faint, high-redshift events, will be observed in the near future by the ground-based detector network. This background is anticipated to exhibit non-Gaussian statistical properties. To develop a robust method for detecting such a non-Gaussian gravitational-wave background, we revisit optimal detection strategies based on the Gaussian-mixture likelihood model. In this work, we demonstrate that properly accounting for the detector antenna pattern is essential. Current approaches typically rely on the overlap reduction function averaged over the sky. Through simulations, we show that using such an averaged response introduces significant biases in parameter estimation. In addition, we propose a computationally feasible method that incorporates second-order corrections as an approximation of the full integral over the source distribution. Our results indicate that this approach effectively eliminates these biases. We also show that our method remains robust even when considering anisotropic backgrounds.
\end{abstract}

\maketitle

\section{Introduction}
Remarkable progress has been made in gravitational wave (GW) detection in recent years. The ground-based detector network operated by the LIGO-Virgo-KAGRA (LVK) collaboration~\cite{LIGOScientific:2014pky, VIRGO:2014yos, KAGRA:2020tym} has successfully detected numerous GW events~\cite{LIGOScientific:2016dsl,LIGOScientific:2018mvr,LIGOScientific:2020ibl,LIGOScientific:2020kqk,LIGOScientific:2021usb,KAGRA:2021vkt,KAGRA:2021duu,LIGOScientific:2025slb,LIGOScientific:2025pvj,LIGOScientific:2025brd}, including binary black holes (BBHs), binary neutron stars, and black hole-neutron star mergers. With planned upgrades to existing detectors and the development of next-generation facilities such as the Einstein Telescope (ET)~\cite{Punturo:2010zz} and Cosmic Explorer (CE)~\cite{Evans:2021gyd}, along with space-based observatories like LISA~\cite{LISA:2024hlh}, TianQin~\cite{TianQin:2015yph}, Taiji~\cite{Ruan:2018tsw}, B-DECIGO and DECIGO~\cite{Kawamura:2020pcg}, we will be able to probe a wider frequency spectrum with significantly enhanced sensitivity. This will allow the detection of more GW signals, including increasingly faint sources. Among these targets, the gravitational wave background (GWB) remains a particularly important focus~\cite{Romano:2016dpx, vanRemortel:2022fkb}.

Stochastic GWB refers to a random and persistent background of GWs. There are many possible sources of GWB signals, including astrophysical sources~\cite{Regimbau:2011rp} such as compact binary coalescing (CBC) systems, supernova explosions~\cite{Buonanno:2004tp}, and cosmological processes~\cite{LIGOScientific:2025kry} such as inflation and phase transitions. Typically, the dimensionless energy density $\Omgw$ is used to describe the strength of the GWB:
\begin{align}
    \Omgw(f) = \frac{1}{\rho_c}\dv{\rho_{\mathrm{gw}}}{\ln{f}} \,,
\end{align}
where $\rho_{\mathrm{gw}}$ is the energy density of the GWB and $\rho_c= 3 H_0^2/(8\pi G)$ is the critical energy density of the Universe today with $H_0$ being the Hubble parameter at present time. The observed individual BBH systems in the LVK O4a run provide insights into the merger rate and redshift distribution of stellar-mass binaries, which can have implications for the amplitude of the astrophysical background. This suggests $\Omgw (25{\rm Hz}) = 0.9^{+1.1}_{-0.5} \times 10^{-9}$, a value that may be detectable with the sensitivity of the O5 run~\cite{LIGOScientific:2016fpe,LIGOScientific:2025bgj}. Measuring such an astrophysical background is essential for understanding the black hole formation history~\cite{Belczynski:2001uc, Bethe:1998bn}.

Detecting the GWB has long been a challenge due to its weak intensity. Achieving a higher signal-to-noise ratio (SNR) requires long-term observational data and cross-correlation between multiple detectors to suppress the influence of noise~\cite{Allen:1997ad}. For the astrophysical background from stellar-mass BBH systems, the average duration of GW signals is shorter than the average interval between two consecutive BBH mergers in the frequency band of ground-based detectors, leading to an intermittent distribution of signals in the data~\cite{Cornish:2015pda}. The duty cycle, $\xi$, quantifies this intermittency by representing the fraction of the total data occupied by GW signals~\cite{Coward:2006df}. A lower $\xi$ implies a higher degree of non-Gaussianity in the strain distribution. 
Based on the observed BBH population, studies suggest that individual signals rarely overlap, yielding an estimated duty cycle of only $\xi \sim 10^{-3}$--$10^{-2}$~\cite{LIGOScientific:2017zlf}. This quantity depends on the BBH population and provides a valuable probe of the origin and population properties of BBHs~\cite{Braglia:2022icu,Bagui:2026hrf}.

We note that, for an astrophysical GWB, the signal consists of a mixture of nearby and distant sources. Since events below the detector sensitivity contribute negligibly unless they overlap sufficiently to produce a detectable stochastic background, the effective duty cycle relevant to a given analysis depends on the detector sensitivity. Consequently, the effective duty cycle may be smaller for the current LVK network than for ET-class detectors because of its lower sensitivity~\cite{Braglia:2022icu}. The duty cycle also depends on the duration of signals in the observing band: long-lived sources, such as neutron-star binaries~\cite{Regimbau:2009rk}, or sources observed at low frequencies by detectors such as LISA or pulsar timing arrays, remain in band for extended periods and are therefore much more likely to overlap, resulting in a high-duty-cycle background. Nevertheless, the amplitude distribution may still exhibit measurable non-Gaussian features~\cite{Buscicchio:2024wwm,Ravi:2012bz,Lamb:2024gbh}.

Therefore, intermittent searches are particularly well suited to astrophysical BBH backgrounds in the LVK/ET/CE frequency band, where the duty cycle is expected to remain low and the signal retains strong non-Gaussian intermittent features.
Drasco and Flanagan~\cite{Drasco:2002yd} were the first to develop a systematic theoretical framework for detecting intermittent GWB. This framework was later generalized to include the conventional cross-correlation model applied to spatially separated interferometers with colored, non-Gaussian noise such as glitches~\cite{Thrane:2013kb}. It has been shown that these non-Gaussian statistics can significantly enhance detection efficiency compared to standard cross-correlation search methods, particularly at low $\xi$~\cite{Lawrence:2023buo}. 

In addition, there have been proposals to utilize higher-order statistics. Theoretically, if the properties of noise and the GW signal are well understood and modeled properly, higher-order statistical measures can enhance detection efficiency~\cite{Buscicchio:2022raf,Ballelli:2022bli}. Seto~\cite{Seto:2008xr,Seto:2009ju} introduced detection methods based on fourth-order statistical quantities and cumulants using four detectors. The work has focused on non-Gaussian statistical properties arising from the intermittency of the GW signal, with the resulting kurtosis parameter being characterized by the duty cycle $\xi$. In contrast, Martellini and Regimbau~\cite{Martellini:2014xia,Martellini:2015mfr} addressed non-Gaussianity in signal distributions and proposed semiparametric maximum likelihood estimators to estimate higher-order moments via a fourth-order Edgeworth expansion, and provided detailed comparisons between second-order and fourth-order statistics. Application of a heavy-tailed Generalized Hyperbolic function has been discussed in~\cite{Sasli:2023mxr,Karnesis:2024pxh}. 

Other approaches also have been extensively developed. One prominent method involves using subthreshold BBHs in matched-filtering searches combined with Bayesian statistics to infer the duty cycle~\cite{Smith:2017vfk,Smith:2020lkj,Biscoveanu:2020gds,Kou:2025bhk}. In addition, spectrogram correlated stacking has been proposed to leverage higher-order statistics in the time-frequency domain~\cite{Dey:2023oui,Sah:2023bgr,Sah:2025agw}. Finally, there has also been new efforts to apply deep learning methods to the search for non-Gaussian backgrounds~\cite{Utina:2021ipo,Yamamoto:2022kuh}.

The non-Gaussian level in the astrophysical background is determined by three factors: {\bf intermittency, sample distribution, and waveform}. Intermittency is particularly significant for the BBH background, which is expected to exhibit an extremely low duty cycle, $\xi \sim {\cal O}(10^{-3})$~\cite{Lawrence:2023buo}. This aspect is typically modeled using a Gaussian mixture framework~\cite{Drasco:2002yd}, where the likelihood comprises two Gaussian components representing segments with and without signals. The segments containing signals are often assumed to have strain amplitudes that follow a Gaussian distribution, based on the central limit theorem. However, this assumption may not hold due to the limited number of samples or when the samples are drawn from a heavy-tailed power-law distribution, which can occur when the sample is taken assuming the typical BBH population. Indeed, recent studies have shown that the statistical properties of the GWB strongly depend on the BBH population, including their mass and redshift distributions~\cite{Sah:2023bgr}. Furthermore, the GW waveform produced by BBH systems exhibits inherent non-Gaussian characteristics. Recent work in the context of pulsar timing~\cite{Lamb:2024gbh} has indicated that the higher-order statistical frequency spectrum of the astrophysical GWB follows a power-law behavior. These properties are closely related to the distinctive dynamical evolution of BBH systems.

In light of these considerations, the analysis method can be enhanced by incorporating additional non-Gaussian aspects, underscoring the importance of accurately modeling such distributional properties. In this paper, we emphasize that, beyond the three factors discussed above, properly accounting for the detector response is crucial for intermittent searches. Because GW detectors have specific non-isotropic antenna patterns, even if the intrinsic GW signal follows a Gaussian distribution, the projected amplitude becomes distorted. 
%We quantify this effect and propose a computationally efficient method to incorporate it. 
We quantify this effect by, for the first time, numerically evaluating the full likelihood that explicitly integrates over the source direction on the sky. We then develop an analytic approximation that accurately captures the effect of the detector response while significantly reducing the computational cost. Furthermore, using simulated data and Bayesian parameter inference, we show that neglecting this effect can lead to significant systematic biases in the inferred duty cycle and GW signal amplitude.

This work is the first in a series of papers. In this paper, we focus on the effect of the antenna pattern function. To present this in a simple framework, we assume intermittent stochastic burst signals, meaning we do not model individual waveforms that compose the stochastic background. We simulate detection using a likelihood based on the Gaussian mixture model~\cite{Drasco:2002yd}, modified to incorporate the modulation introduced by the detector response. In a subsequent paper, we will consider more realistic signals, specifically sub-threshold BBH events, and present a detection method based on a fully non-Gaussian likelihood that accounts for all relevant non-Gaussian factors arising not only from intermittency but also from source sample distribution and waveform characteristics.

The paper is organized as follows: In Sec.~\ref{sec_2}, we review the current detection methods for an intermittent background and introduce basic setups for the simulation. In Sec.~\ref{sec_3}, we analyze the effect of detector response on intermittent searches and propose a computationally efficient method to incorporate this effect. We also examine the impact of anisotropic source distributions. In Sec.~\ref{sec_4}, we present simulation results for signal detection and parameter estimation, comparing different likelihood models and demonstrating how the modified likelihood mitigates parameter bias. We further investigate whether incorrectly assuming an isotropic background when the true distribution is anisotropic introduces additional bias. Finally, Sec.~\ref{sec_5} provides a summary of our findings.

%%%%%%%%%%%%%%%%%%%%%%%%%%%%%%%%%%%%%%%%%%%%%%%%%%%%%%%%%%%
%%%%%%%%%%%%%%%%%%%%%%%%%%%%%%%%%%%%%%%%%%%%%%%%%%%%%%%%%%%
\section{Intermittent search and cross-correlation}\label{sec_2}

\subsection{The Gaussian mixture model}\label{subsec_2_1}
Intermittency is typically characterized by a quantity called the duty cycle $\xi$, which ranges from 0 to 1 and represents the fraction of data segments containing a GW signal~\cite{Drasco:2002yd}. %More specifically, if the observational data over a given period is divided into $K$ segments, the duty cycle can be approximated as, $\xi \approx \Ngw/K$ where $\Ngw$ is the number of segments that contain GW signals and $K$ is the total number of segments. Typically, time-series data is divided into short segments optimized for the target signal. 
Long time-series data (months or years) are typically divided into segments. Using the total number of segments $N_{\rm seg}$ and the number of segments that contain GW signals $\Ngw$, the duty cycle is defined as $\xi := \Ngw/N_{\rm seg}$. The segment length is typically optimized for the target signal. 
For the BBH background, a segment length of about 4 seconds is commonly chosen to ensure that a single signal (with a duration of $\sim 1$s) is contained within one segment, while still being long enough to provide a stable noise estimate. 

It is important to note that the definition of such data-driven duty cycle involves some ambiguity. Recently, a potential bias in parameter estimation caused by segment size was highlighted in~\cite{Franciolini:2025leq}. Furthermore, it depends on the detector sensitivity, as events with extremely low SNR ($\ll 1$), which are typically numerous, do not contribute to the analysis~\cite{Braglia:2022icu}. Naively, one might set the threshold for inclusion in the analysis based on the matched-filtering SNR of each event to be larger than $1$; however, it remains unclear whether this is an appropriate or optimal choice for the threshold. 

In contrast, the duty cycle can also be calculated theoretically by counting the number of events that constitute the GWB and accounting for the duration of each event~\cite{Coward:2006df,Rosado:2011kv,Braglia:2022icu}. Establishing a robust connection between the theoretical and data-driven duty cycles remains a challenge due to these ambiguities. This is essential for extracting astrophysical information from the data, but is beyond the scope of this paper. 
%In this work, we generate mock data by injecting Gaussian bursts, ensuring that each signal-containing segment is occupied by a burst-like signal whose amplitude is drawn from a Gaussian distribution. Therefore, we do not expect this ambiguity to arise.

Drasco and Flanagan~\cite{Drasco:2002yd} were the first to propose using a Gaussian mixture model to develop detection statistics optimized for non-Gaussian backgrounds. 
The intermittent background is modeled by assuming that each data segment belongs to one of two possible states: a signal-present state and a signal-absent state. Let ${\cal H}_1$ denote the hypothesis that a GW signal is present in the segment and ${\cal H}_0$ the hypothesis that no GW signal is present, and $\Like_s(\pmb{s}|{\cal H}_0)$ and $\Like_n(\pmb{s}|{\cal H}_1)$ denote the likelihood functions for the data $\pmb{s}$, respectively. 

Consider a stochastic background generated by a population of discrete burst events. Let $N$ denote the expected number of signals present in a data segment. Assuming that event occurrences follow a Poisson process, the probability of finding $n$ signals in a segment is 
\begin{equation} 
P(n)=\frac{N^n e^{-N}}{n!}. 
\end{equation} 
The probability that a segment contains at least one signal is therefore \begin{equation} 
\xi = 1-P(0)=1-e^{-N}. 
\end{equation} 
For the intermittent backgrounds considered in this work we focus on the regime $\xi\ll1$, corresponding to $N\ll1$. In this limit, 
\begin{equation} 
\xi = 1-e^{-N} \simeq N, 
\end{equation} 
and most segments contain either zero or one signal. Neglecting the probability of multiple signals in a segment, the data can be described by two mutually exclusive hypotheses:
\begin{equation} 
p_{\rm tot}({\bf s}) = \xi\, p({\bf s}|H_1) + (1-\xi)\, p({\bf s}|H_0). 
\label{eq:ptot}
\end{equation} 
Assuming statistical independence between segments, the likelihood for the full dataset is therefore 
\begin{align}
\Like_{\tot}(\pmb{s})=\prod_i \Big[ \xi\Like_s(\pmb{s}_i|{\cal H}_1) + (1-\xi)\Like_n(\pmb{s}_i|{\cal H}_0) \Big] \,,
\label{eq:likelihood_total}
\end{align}
where $i$ denotes the index of each data segment.

Eq.~\eqref{eq:ptot} can equivalently be interpreted as a mixture model for the GW strain amplitude $h$~\cite{Drasco:2002yd}, 
\begin{align} 
p_{\rm tot}(h) = \xi p_h(h) + (1-\xi)\delta(h) \,, 
\end{align} 
where $p_h(h)$ is the strain-amplitude distribution of individual GW events and $\delta(h)$ represents segments containing no signal. 
%In this framework, the overall amplitude of the GWB is characterized by the variance of the distribution $p_{\rm tot}(h)$.

The distribution $p_h(h)$ is often assumed to be Gaussian, primarily because, in an ideal scenario with a large volume of observational data, the dataset would contain multiple independent GW signals. Under these conditions, $p_h(h)$ would closely approximate a Gaussian distribution due to the central limit theorem. In the context of intermittent searches for the BBH background, however, this assumption may not hold, because the BBH source population gives rise to a highly skewed distribution dominated by many weak events and only a few strong nearby sources. Nevertheless, in this paper, we focus on the effect of the antenna pattern function and therefore assume a Gaussian distribution for the original signal amplitudes. This choice simplifies the analysis and allows us to clearly isolate and study the impact of the antenna pattern. 
The non-Gaussianity of $p_h(h)$ itself, and its coupling to the non-Gaussianity induced by the antenna pattern functions, will be investigated in our future studies.

\subsection{Cross-correlating detectors in different locations}\label{subsec_2_2}

In this work, we relax the assumption of co-aligned and co-located detectors adopted in previous studies~\cite{Drasco:2002yd,Lawrence:2023buo}. 
By doing so, we are required to consider the response of cross-correlated detectors, which is often described by the overlap reduction function~\cite{Allen:1997ad}. One of the key points of this paper is that accounting for the antenna pattern function introduces non-trivial effects on the signal distribution, as the response is not spherically symmetric and alters the distribution of the projected GW strain amplitude. We describe the details of our setup below.

For ground-based detectors, where the light travel time along the arms is short compared to the GW period, the antenna pattern function in the small-antenna limit can be approximated as
\begin{align}
    F_P(\hatOmg) = \frac{1}{2}e^{ab}_P (\hatOmg) \left(\hat{l}^a_1 \hat{l}^b_1 - \hat{l}^a_2 \hat{l}^b_2\right) \,,
\end{align}
where $\hatOmg$ is the direction in which the GWs propagate, $\hat{l}_{1,2}$ are the unit vectors describing the two arms of the detector, and $e_P^{ab}$ with $P=+,\times$ are transverse-traceless polarization tensors,
\begin{align}
    &e_+^{ab}(\hatOmg) = \hat{p}^a\hat{p}^b-\hat{q}^a\hat{q}^b, \\
    &e_\times^{ab}(\hatOmg) = \hat{p}^a\hat{q}^b + \hat{q}^a\hat{p}^b .
\end{align}
Here, $\hat{p}$ and $\hat{q}$ are vectors perpendicular to each other and to $\hatOmg$. Considering the earth's rotation, we have~\cite{Jaranowski:1998qm}
\begin{align}
    &F_+(\hatOmg) = \sin{\zeta}\big( \cos{2\psi} G_+(\hatOmg) + \sin{2\psi}G_\times(\hatOmg) \big), \\
    &F_\times(\hatOmg) = \sin{\zeta}\big( \cos{2\psi} G_\times(\hatOmg) - \sin{2\psi} G_+(\hatOmg) \big) \,,
\end{align}
and $G_{+,\times}$ are given by
\begin{align}
    &G_+(\hatOmg) = \frac{1}{4}\cos{(\alpha-\varphi-\omega_{\oplus}t)}\sin{2\beta}\sin{2\delta}\sin{2\lambda}\nonumber\\
    & + \frac{1}{16}(3-\cos{2\delta})(3-\cos{2\lambda})\cos{2(\alpha-\varphi-\omega_{\oplus}t)}\sin{2\beta} \nonumber\\
    & - \frac{1}{2}\cos{2\beta}\cos{\lambda}\sin{2\delta}\sin{(\alpha-\varphi-\omega_{\oplus}t)} \nonumber \\
    & - \frac{1}{4}\cos{2\beta}(3-\cos{2\delta})\sin{\lambda}\sin{2(\alpha-\varphi-\omega_{\oplus}t)} \nonumber\\
    & + \frac{3}{4}\cos^2{\delta}\cos^2{\lambda}\sin{2\beta}, \\
    &G_\times (\hatOmg) = \cos{2\beta}\cos{\delta}\cos{\lambda}\cos{(\alpha-\varphi-\omega_{\oplus}t)} + \nonumber\\
    & + \cos{2\beta}\cos{2(\alpha-\varphi-\omega_{\oplus}t)}\sin{\delta}\sin{\lambda} \nonumber\\
    & + \frac{1}{2}\cos{\delta}\sin{2\beta}\sin{2\lambda}\sin{(\alpha-\varphi-\omega_{\oplus}t)} \nonumber\\
    & + \frac{1}{4}(3-\cos{2\lambda})\sin{2\beta}\sin{\delta}\sin{2(\alpha-\varphi-\omega_{\oplus}t)},
\end{align}
where $(\zeta, \lambda, \varphi, \beta)$ are parameters characterizing the detector: $\zeta$ is the angle between the detector's arms, $\lambda$ and $\varphi$ are the detector's latitude and longitude, respectively, and $\beta$ is the angle between the arm bisector and the East direction. The parameters $(\alpha, \delta)$ represent the right ascension and declination of the source direction, corresponding to the direction $-\hatOmg$, and $\omega_{\oplus}$ is the Earth's rotational angular velocity; for simplicity, this term is omitted.

As convention, we consider the decomposition of the tensor metric perturbations $h_{ab}$ into the Fourier modes using the two independent polarization states
\begin{equation}
 h_{ab} (t,\textbf{x}) = \sum_P \int^{\infty}_{-\infty} \dd f
 \int_{S^2} \dd\hatOmg \,
 \Hf_P(f,\hatOmg) \ee^{\ii2\pi f(t-\hatOmg\cdot \textbf{x})}e_{ab}^P(\hatOmg) \,, 
 \label{eq:h_ab}
\end{equation}
where $\textbf{x}$ describes the location of the detector and $\Hf_P(f,\hatOmg)$ is the amplitude of each polarization mode. 
Now, even though our interest lies in an isotropic (intermittent) background, we will show that the directionality of the background in each segment must still be taken into account. To demonstrate this explicitly, we retain the anisotropy of the background in the formulation in order to examine the effect of the antenna pattern function.  For a stationary, Gaussian, and unpolarized background, the amplitudes satisfy the following statistical
properties,
\begin{eqnarray}
&& \langle 
\Hf^*_P(f,\hatOmg) \Hf_{P'}(f',\hatOmg') 
\rangle \nonumber \\
&& ~~~ = \frac{1}{4} \delta_{P,P'} \delta(f-f') \frac{1}{4\pi}\delta^2(\hatOmg,\hatOmg')  {\cal P}_H(f,\hatOmg) \,,
\label{eq:power_definition}
\end{eqnarray}
where the power spectral density (PSD) ${\cal P}_H(f,\hatOmg)$ characterizes both the spectral and angular distribution of the background. 
Assuming that the anisotropy is frequency‑independent, the PSD can be factorized as 
${\cal P}_H(f,\hatOmg)=P_H(f){\cal P}(\hatOmg)$ with $\int \dd \hatOmg {\cal P}(\hatOmg) = 4\pi$. The factor $1/4$ is included by convention to account for the one-sided definition of the strain power spectrum and summing the two polarization states, while the $1/4\pi$ factor reflects the normalization over the unit sphere\footnote{Our definition differs from that used in anisotropic background searches (see e.g.,~\cite{Thrane:2009fp}) by a factor of $1/4\pi$. This is because, in our case, the GWB amplitude of interest remains isotropic after summing over contributions from all directions. In contrast, anisotropic searches focus on continuous sources with directional dependence, where the signal is localized or enhanced in a specific direction.}.
Then it is related to the GW energy density parameter as 
\begin{equation}
\Omgw (f) = \frac{2\pi^2}{3H_0^2}f^3 \frac{1}{4\pi} \int \dd \hatOmg ~ {\cal P}_H(f,\hatOmg) \,. 
\label{eq:def_gwb_density}
\end{equation}
%Note that, for an intermittent background, the energy density should be multiplied by the duty cycle $\xi$, which represents the fraction of time during which the signal is present, effectively reducing the observed energy density compared to a continuous background.

Note that the duty cycle also affects the observed GW energy density. The quantity $\Omgw$ is defined from the GW energy density averaged over the observation time. For an intermittent background, GWs are present only during a fraction $\xi$ of the observing time. If the GW power spectrum in signal-present segments is denoted by $P_H(f)$, the time-averaged spectrum becomes 
\begin{equation} 
\langle P_H(f) \rangle = \xi P_H(f), 
\end{equation} 
since the remaining fraction $1-\xi$ contributes no GW power. Using Eq.~\eqref{eq:def_gwb_density}, the corresponding observed energy density of an intermittent background, denoted by $\Omgw^{\rm int}$, is 
\begin{equation} 
\Omgw^{\rm int}(f) = \frac{2\pi^2}{3H_0^2}f^3 \frac{1}{4\pi} \int \dd\hatOmg\, \langle {\cal P}_H(f,\hatOmg) \rangle = \xi \Omgw(f). 
\end{equation} 
Therefore, the duty cycle acts as a suppression factor for the time-averaged GW energy density.

Using the antenna pattern function, the GW signal measured at the detector $I$ in the Fourier space is given by
\begin{align}
\Hf_I(f) = \sum_P \int_{S^2} \dd \hatOmg  \Hf_P(f,\hatOmg) F_{I,P}(\hatOmg) \ee^{-\ii 2\pi f \hatOmg \cdot {\bf x}_I} \,.
\label{eq:gw_in_detectorI}
\end{align}
In GWB searches with ground-based detectors, the data is divided into short time segments and analyzed in the discrete Fourier domain. For a segment that contains a GW signal, the observable $\sf$ in detector $I$ in the $i$-th data segment can be expressed as:
\begin{align}
\sf_{I,i}(f) = \nf_{I,i}(f) + \Hf_{I,i}(f) \,,
\end{align}
where $\nf_{I,i}$ represents the detector noise in Fourier space. 

When performing cross-correlation between two detectors, the noise terms are suppressed under the assumption that the noise is uncorrelated, and the expectation value of the cross-correlated data can be approximated as
\begin{eqnarray}
&&\langle \sf^*_{I,i}(f)\sf_{J,i}(f') \rangle \simeq
\langle \Hf^*_{I,i}(f)\Hf_{J,i}(f') \rangle
\nonumber\\
&& ~~~ =
\sum_{P,P'} \int_{S^2} \dd \hatOmg 
\int_{S^2} \dd \hatOmg' 
\langle \Hf^*_P(f,\hatOmg)\Hf_{P'}(f',\hatOmg') \rangle
\nonumber\\
&& ~~~~~~~~~~~~~~~~ \times \, F_{I,P}(\hatOmg)F_{J,P'}(\hatOmg')
\ee^{\ii 2\pi f \hatOmg \cdot (\pmb{\mathrm{x}}_I - \pmb{\mathrm{x}}_J)} \,.
\label{Eq:ss}
\end{eqnarray}

For an isotropic background, $\Hf_P(f,\hatOmg)$ becomes independent of direction and we can set ${\cal P}(\hatOmg)=1$. In this case, by substituting Eq.~\eqref{eq:power_definition}, Eq.~\eqref{Eq:ss} simplifies to~\cite{Allen:1997ad}
\begin{equation}
\langle \sf^*_{I,i}(f)\sf_{J,i}(f') \rangle \simeq \frac{1}{10} \delta(f-f')  P_H(f) \bar\gamma_{IJ}(f) \,,
\end{equation}
where the overlap reduction function is defined as 
\begin{equation}
\bar\gamma_{IJ}(f) := \frac{5}{8\pi}\sum_P \int_{S^2} \dd \hatOmg \, F_{I,P}(\hatOmg)F_{J,P}(\hatOmg)
\ee^{\ii 2\pi f \hatOmg \cdot (\pmb{\mathrm{x}}_I - \pmb{\mathrm{x}}_J)} \,.
\end{equation}
In the case of the intermittent astrophysical background, the signal is not isotropic in each segment. Instead, it behaves like a point source, which can be modeled as: ${\cal P}(\hatOmg)=4\pi\delta^2(\hatOmg,\hatOmg_i)$ with $\hatOmg_i$ representing the GW direction in $i$-th data segment. By explicitly leaving the directional dependence, Eq.~\eqref{Eq:ss} leads to
\begin{equation}
\langle \sf^*_{I,i}(f)\sf_{J,i}(f') \rangle \simeq
\frac{1}{4} \delta(f-f')  P_H(f) \gamma_{IJ}(f, \hatOmg_i) \,,
\end{equation}
where
\begin{equation}
\gamma_{IJ}(f, \hatOmg_i) := \sum_P F_{I,P}(\hatOmg_i)F_{J,P}(\hatOmg_i)
\ee^{\ii 2\pi f \hatOmg_i \cdot (\pmb{\mathrm{x}}_I - \pmb{\mathrm{x}}_J)} \,.
\label{eq:gammaIJ}
\end{equation}
We typically assume a uniform distribution of astrophysical sources over the sky, and take the ensemble average over source directions which is denoted as $\langle \gamma_{IJ}(f, \hatOmg_i) \rangle_{\hatOmg_i}$. This averaging leads to $\langle \gamma_{IJ}(f, \hatOmg_i) \rangle_{\hatOmg_i} = 2 \bar\gamma_{IJ}(f)/5$ and it becomes effectively equivalent to the isotropic case~\cite{Allen:1996gp}.

\subsection{Likelihood function}
%Let us assume that the noise is stationary, Gaussian-distributed, and that its real and imaginary components in the Fourier domain, as well as different frequency bins, are statistically independent. 
We assume that the detector noise is stationary and Gaussian distributed. Under the stationarity assumption, different frequency bins in the Fourier domain are uncorrelated, resulting in a diagonal covariance matrix. Furthermore, for Gaussian noise, uncorrelated Fourier components are statistically independent. The real and imaginary parts of each Fourier mode are also independent and have equal variance.
Under these assumptions, the probability density function of the noise can be expressed as
\begin{align}
p_{n_I}(\nf_I(f))=\Gauss[\sigma^2_{n_I}(f)](\Re\nf_I(f))\Gauss[\sigma^2_{n_I}(f)](\Im\nf_I(f)),
\end{align}
where $\Gauss[\sigma^2](x):=e^{-x^2/(2\sigma^2)}/\sqrt{2\pi\sigma^2}$ denotes the standard Gaussian distribution function. 
Here, we explicitly separate the real and imaginary parts of the noise. The variance $\sigma^2_{n_I}(f)$ is defined as
\begin{align}
\sigma^2_{n_I}(f):=\frac{\Tseg}{4}P_{n_I}(f),
\end{align}
where $\Tseg$ is the duration of the data segment, and $P_{n_I}(f)$ is the noise PSD of detector $I$. 

So far, we have expressed quantities as continuous functions of frequency. However, in practice, we work with discrete frequency bins defined by $f_k=k\Delta f$ with $\Delta f = 1/T_{\rm seg}$ being the frequency resolution determined by the segment duration $T_{\rm seg}$. The likelihood function is then constructed as a product over the likelihoods of individual frequency bins. For a signal-absent segment, the likelihood function $\Like_n(\pmb{s}_i|{\cal H}_0)$ is given by
\begin{equation}
\Like_n(\sf_i|\sigma^2_n) = 
\prod_k \prod_I \frac{1}{2\pi\sigma^2_{n_I}(f_k)} \exp\left[-\frac{1}{2}\frac{\sf_{I,i}^*(f_k)\sf_{I,i}(f_k)}{\sigma^2_{n_I}(f_k)}\right]\,.
\label{eq:likelihood_n}
\end{equation}

Assuming that the GW signal $\Hf_P$ follows a Gaussian distribution, its probability density function is given by:
\begin{align}
p_{\gw}(\Hf_P(f))=
\Gauss[\sigma^2_H(f)](\Re\Hf_P(f))
\Gauss[\sigma^2_H(f)](\Im\Hf_P(f)),
\end{align}
where 
\begin{equation}
\sigma^2_H(f):= \frac{T_{\rm seg}}{2} P_H(f),
\label{eq:sigmaH2}
\end{equation}
representing the variance of the GW signal in the Fourier domain. This variance relates to the intensity of the GWB that we aim to detect, with $\Omgw^{\rm int}(f) \propto \xi \sigma^2_H(f)$.

Accordingly, we can construct the likelihood function of the signal segment $\Like_s(\pmb{s}_i|{\cal H}_1)$ as
\begin{align}
&\Like_s(\sf_i|\sigma^2_n,\sigma^2_H)\nonumber\\
&=\prod_k \int \prod_I p_{n_I}\bigg(\sf_{I,i}
(f_k)
- \sum_{P} F_{I,P}(\hatOmg_i) \ee^{-\ii 2\pi f_k \hatOmg_i \cdot {\bf x}_I} \nonumber\\
&\quad\times \Hf_P(f_k) \bigg)p_\Omega(\hatOmg_i)\dd{\hatOmg_i} \prod_{P} p_{\gw}(\Hf_P(f_k))\dd{\Hf_P(f_k)}. \label{eq:likelihood_s_int}
\end{align}
Here, $p_\Omega(\hatOmg_i)$ is the probability density function of GW direction $\hatOmg_i$. Substituting the Gaussian distributions and performing the integral, one would obtain~\cite{Cornish:2015pda,Matas:2020roi}:
\begin{align}
&\Like_s(\sf_i|\sigma^2_n,\sigma^2_H)=\prod_k \int \frac{1}{\det[2\pi\Cov(f_k,\hatOmg_i)]}\nonumber\\
&\times \exp\left[-\frac{1}{2}\sum_{IJ}\sf_{I,i}^*
(f_k)
(\Cov^{-1}(f_k,\hatOmg_i))_{IJ}\sf_{J,i}
(f_k)
\right] p_\Omega(\hatOmg_i)\dd{\hatOmg_i},
\label{eq:likelihood_s}
\end{align}
where $\Cov$ is the covariance matrix, given by:
\begin{widetext}
\begin{align}
&\Cov(f_k,\hatOmg_i):=\left[
\begin{array}{ccc}
\sigma^2_{n_1}(f_k)+|\gamma_1|^2(\hatOmg_i)\sigma^2_H(f_k) & \gamma_{12}^*(f_k,\hatOmg_i)\sigma^2_H(f_k) & ... \\
\gamma_{12}(f_k,\hatOmg_i)\sigma^2_H(f_k) & \sigma^2_{n_2}(f_k)+|\gamma_2|^2(\hatOmg_i)\sigma^2_H(f_k) & ... \\
... & ... & ...
\end{array}\right] \,,
\label{eq:cov}
\end{align}
\end{widetext}
where $\gamma_{IJ}$ is the non-averaged overlap reduction function defined in Eq.~\eqref{eq:gammaIJ} and we define
\begin{equation}
    |\gamma_I|^2(\hatOmg_i):=\sum_P |F_{I,P}(\hatOmg_i)|^2 .
\end{equation}
The full derivation is presented in Appendix~\ref{app:deriv_likelihood_s}.

In practical detection scenarios, evaluating the integral over $\dd \hatOmg$ is computationally challenging. 
In the following sections, we estimate this integral numerically using a Monte Carlo method. For a given angular distribution $p_\Omega$ and detector configuration, we generate samples of the antenna pattern functions, evaluate the integrand in Eq.~\eqref{eq:cov} for each sample, and approximate the integral by the sample average. We use $1000$ samples for each evaluation, which is not particularly large but represents a reasonable compromise between numerical accuracy and computational cost for the purposes of the present study. While the calculation is feasible in our simplified simulation setup, which assumes a short observation time and does not simultaneously estimate the noise properties, the computational cost is already substantial and would likely become prohibitive in a realistic search scenario.

One approach commonly adopted in previous studies to simplify the analysis 
and achieve a reasonable computation time 
is to assume a uniform prior $p_\Omega$ over the sky and replace $\gamma_{IJ}(f, \hatOmg_i)$ and $|\gamma_I|^2(\hatOmg_i)$ with their directional averages $\langle \gamma_{IJ}(f, \hatOmg_i) \rangle_{\hatOmg_i} = 2 \bar\gamma_{IJ}(f)/5$ and $\langle |\gamma_I|^2(\hatOmg_i) \rangle_{\hatOmg_i} = 2/5$, respectively. This approximation removes the $\hatOmg_i$-dependence from the covariance matrix and simplifies the calculation. We denote this direction-averaged covariance matrix as $\avg{\Cov}_{\hatOmg_i}$ for later use, 
\begin{align}
&\avg{\Cov}_{\hatOmg_i}(f_k):=\nonumber\\
&\left[
\begin{array}{ccc}
\sigma^2_{n_1}(f_k)+\frac{2}{5}\sigma^2_H(f_k) & \frac{2}{5}\bar\gamma_{12}(f_k)\sigma^2_H(f_k) & ... \\
\frac{2}{5}\bar\gamma_{12}(f_k)\sigma^2_H(f_k) & \sigma^2_{n_2}(f_k)+\frac{2}{5}\sigma^2_H(f_k) & ... \\
... & ... & ...
\end{array}\right] \,.
\label{eq:cov_ave}
\end{align}
However, as we will discuss in the following section, this simplification introduces certain limitations.

%%%%%%%%%%%%%%%%%%%%%%%%%%%%%%%%%%%%%%%%%%%%%%%%%%%%%%%%%%%
%%%%%%%%%%%%%%%%%%%%%%%%%%%%%%%%%%%%%%%%%%%%%%%%%%%%%%%%%%%
\section{Effect of detector response}\label{sec_3}
The intermittent search is performed to maximize the likelihood function, given in Eq.~\eqref{eq:likelihood_total}. The noise-related and signal-related likelihood terms, $\Like_n(\pmb{s}_i | {\cal H}_0)$ and $\Like_s(\pmb{s}_i | {\cal H}_1)$, are now given by Eq.~\eqref{eq:likelihood_n} and  Eq.~\eqref{eq:likelihood_s}, respectively. In this section, we discuss how using the direction-averaged covariance matrix $\avg{\Cov}_{\hatOmg_i}$, given in Eq.~\eqref{eq:cov_ave}, instead of performing integration over $\hatOmg_i$ in Eq.~\eqref{eq:likelihood_s}, introduces biases in parameter estimation.

\subsection{Accounting for antenna pattern as second-order corrections}\label{subsec_3_1}

%%%%%%%%%%%%%%%%%%%%%%%%%%%%%%%%%%%%%%%%%%%%%%%%%%%%%%%%
\begin{figure}
\begin{tabular}{c}
\includegraphics[width=0.5\textwidth]{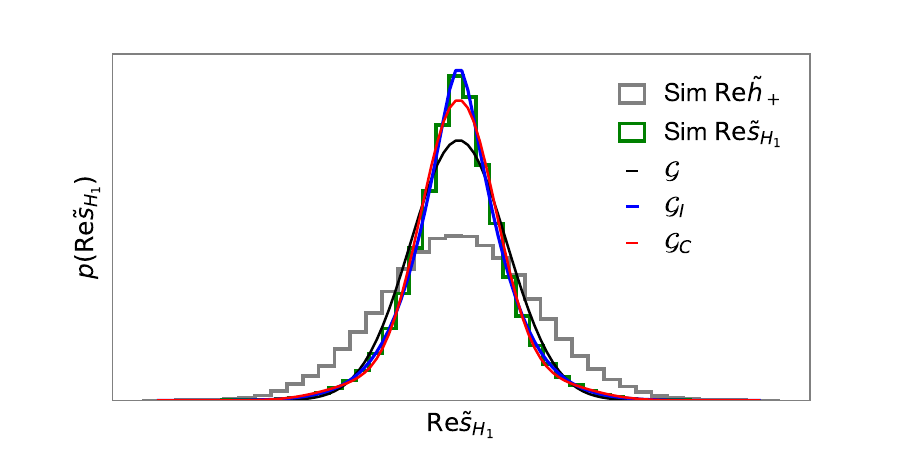}
\end{tabular}
\caption{Schematic representation of the deformation in the GWB distribution after projection onto the detector $H_1$. The grey and green histograms show the distribution of $\Re\hf_+$ (Gaussian) and the received signal by the detector $\Re\sf_{H_1}$, respectively. Solid curves indicate three different theoretical predictions: the blue curve ($\mathcal{G}_I$) shows the fully integrated model calculated as in Eq.~(\ref{eq:cov}), the red curve ($\mathcal{G}_C$) depicts the Gaussian distribution corrected by the second-order covariance matrix $\Gamma_{ab}$ as in Eq.~(\ref{eq_Taylor_L}), and the black curve ($\mathcal{G}$) represents the leading-order Gaussian approximation.}
\label{fig_sim_compared_dist}
\end{figure}
%%%%%%%%%%%%%%%%%%%%%%%%%%%%%%%%%%%%%%%%%%%%%%%%%%%%%%%%

%%%%%%%%%%%%%%%%%%%%%%%%%%%%%%%%%%%%%%%%%%%%%%%%%%%%%%%%
\begin{figure*}[t]
\begin{tabular}{cc}
\includegraphics[width=0.5\textwidth]{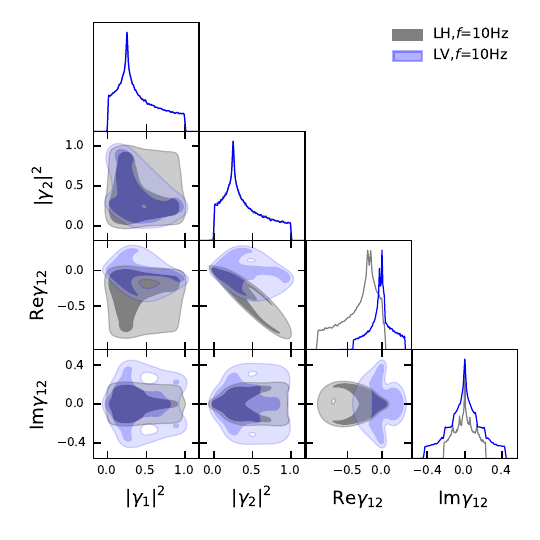}
\includegraphics[width=0.5\textwidth]{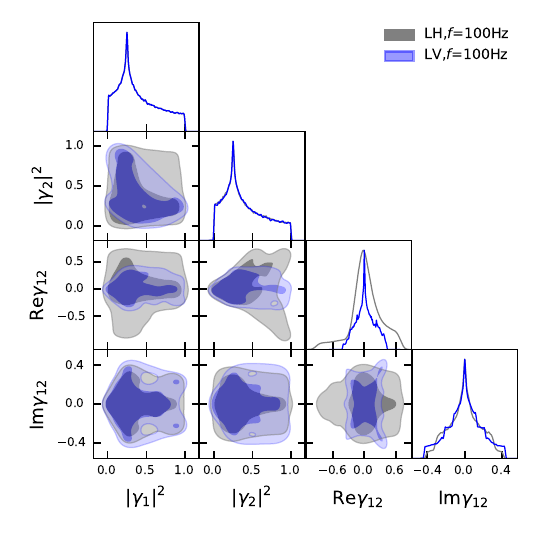}
\end{tabular}
\caption{Distributions of the quantities related to the GW detector projection vector $\vec{y}=(|\gamma_1|^2,|\gamma_2|^2,\Re\gamma_{12},\Im\gamma_{12})$ for isotropic source distribution. The gray color represents the detector pair LIGO-Livingston vs. LIGO-Hanford, while the blue color represents the detector pair LIGO-Livingston vs. Virgo. The left and right panels correspond to $10$Hz and $100$Hz, respectively.}
\label{fig_trig_gamma}
\end{figure*}
%%%%%%%%%%%%%%%%%%%%%%%%%%%%%%%%%%%%%%%%%%%%%%%%%%%%%%%%

%%%%%%%%%%%%%%%%%%%%%%%%%%%%%%%%%%%%%%%%%%%%%%%%%%%%%%%%
\begin{figure*}[t]
\begin{tabular}{c}
\includegraphics[width=\textwidth]{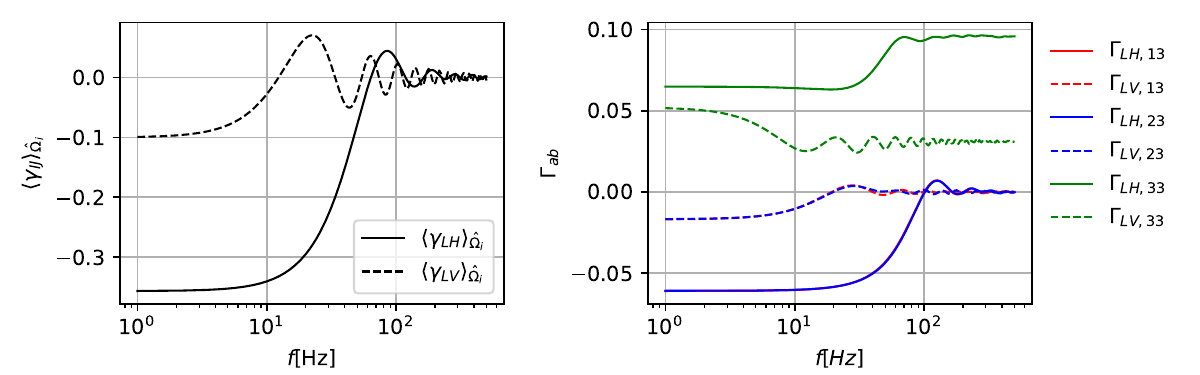}
\end{tabular}
\caption{Frequency dependence of the mean and the second-order covariance matrix elements of $\vec{y}$ for an isotropic source distribution. Left panel: The sky-averaged overlap reduction function of $\langle \gamma_{IJ}(f, \hatOmg_i) \rangle_{\hatOmg_i}$, which is related to the standard overlap reduction function by $\langle \gamma_{IJ}(f, \hatOmg_i) \rangle_{\hatOmg_i} = 2 \bar\gamma_{12}(f)/5$. Right panel: The second-order covariance matrix elements $\Gamma_{ab}$, with different colors representing different components. Solid lines correspond to the LIGO-Livingston vs. LIGO-Hanford detector pair, while dashed lines represent the LIGO-Livingston vs. Virgo pair. Note that the lines corresponding to the (1,3) and (2,3) elements are overlapping.}
\label{fig_corr_compare}
\end{figure*}
%%%%%%%%%%%%%%%%%%%%%%%%%%%%%%%%%%%%%%%%%%%%%%%%%%%%%%%%

The cause of the discrepancy between using  $\avg{\Cov}_{\hatOmg_i}$ and ${\Cov}_{\hatOmg_i}$ may be interpreted as follows. When employing the direction-averaged covariance matrix $\avg{\Cov}_{\hatOmg_i}$, the likelihood function is effectively equivalent to assuming that the observed GW signal at detector $I$, $\Hf_{I,P}(f_k) = F_{I,P}(\hatOmg_i)e^{-\ii 2\pi f_k \hatOmg_i \cdot {\bf x}_I} \Hf_P(f_k)$ follows a Gaussian distribution. However, in reality, due to the non-trivial structure of the antenna pattern function, the projected signal $\Hf_{I,P}(f_k)$ exhibits deviations from the Gaussian assumption, even if the original signal itself is Gaussian. By performing the integration in Eq.~\eqref{eq:likelihood_s}, we can properly account for the deformation of the distribution introduced by the antenna pattern. In contrast, when using the direction-averaged covariance matrix $\avg{\Cov}_{\hatOmg_i}$, as in Eq.~\eqref{eq:cov_ave}, the assumed likelihood shape deviates from the true distribution and eventually introduce bias in the parameter estimation. 

To account for the effect of the antenna patterns, the integration of Eq.~\eqref{eq:likelihood_s} must be evaluated numerically. However, this process is computationally expensive. Instead, we propose the Taylor expanded form:
\begin{align}
\Like_s(\sf_i|\sigma^2_n,\sigma^2_H)&=
\langle\Like_s(\sf_i|\Cov(\hatOmg_i))\rangle_{\hatOmg_i}
\nonumber\\
&\simeq \Like_s(\sf|\avg{\Cov}_{\hatOmg_i})
+\frac{1}{2}\partial_a\partial_b\Like_s(\sf_i|\avg{\Cov}_{\hatOmg_i})\Gamma_{ab} \nonumber\\
&\qquad +\order{\langle y-\langle y\rangle_{\hatOmg_i}\rangle_{\hatOmg_i}^3},
\label{eq_Taylor_L}
\end{align}
where 
\begin{align}
&\Like_s(\sf_i|\Cov(\hatOmg_i)):=\prod_k \frac{1}{\det[2\pi\Cov(f_k,\hatOmg_i)]}\nonumber\\
&\qquad\times \exp\left[-\frac{1}{2}\sum_{IJ}\sf_{I,i}^*(f_k)(\Cov^{-1}(f_k,\hatOmg_i))_{IJ}\sf_{J,i}(f_k)\right] \,.
\end{align}
The partial derivatives of $\Like_s$ are taken with respect to the vector $\vec{y} = (|\gamma_1|^2, |\gamma_2|^2, \Re\gamma_{12}, \Im\gamma_{12}, \ldots)$, and the covariance matrix $\Gamma_{ab} := \langle(y^a - \langle y^a\rangle_{\hatOmg_i})(y^b - \langle y^b\rangle_{\hatOmg_i})\rangle_{\hatOmg_i}$ characterizes the second-order statistical properties of $\vec{y}$. The first term in the expansion corresponds to the Gaussian component, namely the likelihood function with the direction-averaged covariance matrix, which represents the leading-order approximation of the full integral over the angular distribution. 

Eq.~(\ref{eq_Taylor_L}) is equivalent to the Edgeworth expansion~\cite{Feller1971}. This higher-order correction suggests that the likelihood function is no longer Gaussian.
The correction term $\partial_a\partial_b\Like_s(\sf_i|\Cov)$ takes the form
\begin{align}
    &\frac{1}{2}\partial_a\partial_b \Like_s(\sf_i|\Cov) = -\frac{1}{2}\Big( \Sigma^{ab} + \sf_{I,i}\sf_{J,i}^*\Sigma_{IJ}^{ab} \nonumber\\
    &\quad + \sf_{I,i}\sf_{J,i}^*\sf_{K,i}\sf_{L,i}^*\Sigma_{IJKL}^{ab} \Big)\Like_s(\sf_i|\avg{\Cov}_{\hatOmg_i}),
\end{align}
where
\begin{align}
    &\Sigma^{ab} := \Cov^{-1}_{\nu\mu}\partial_a\partial_b\Cov_{\mu\nu} - 2\Cov^{-1}_{\tau(\mu}\Cov^{-1}_{\underline{\nu}\rho)}\partial_a\Cov_{\mu\nu}\partial_b\Cov_{\rho\tau}, \\
    &\Sigma_{IJ}^{ab} := 2\Cov^{-1}_{I\mu}\Cov^{-1}_{(\nu\underline\rho}\Cov^{-1}_{\tau) J} \partial_{(a}\Cov_{\underline\mu\underline\nu}\partial_{b)}\Cov_{\rho\tau} \nonumber\\
    &\quad - \Cov^{-1}_{I\mu}\Cov^{-1}_{\nu J}\partial_a\partial_b\Cov_{\mu\nu}, \\
    &\Sigma_{IJKL}^{ab} := -\frac{1}{4}\Cov^{-1}_{I\mu}\Cov^{-1}_{\nu J}\Cov^{-1}_{K\rho}\Cov^{-1}_{\tau L} \partial_a\Cov_{\mu\nu}\partial_b\Cov_{\rho\tau}.
\end{align}
The parentheses in the subscript denote the symmetry operation applied to the indices. The underlined symbol within the parentheses signifies that no symmetry operation is applied to it.

This correction causes the detector signal distribution to become steeper. Specifically, this is manifested in the fourth-order cumulant of the $\Re\sf_{I,i}$ distribution, i.e., the Kurtosis, becoming positive,
\begin{align}
    &\Var[\Re\sf_{I,i}]\big|_{\sigma^2_{n_{I,i}}=0} := \int (\Re\sf_{I,i})^2 p(\sf_i|\sigma^2_{n_{I,i}}=0,\sigma^2_H)\dd\sf_{I,i} \nonumber\\
    & \quad = \avg{\Cov_{II}}_{\hatOmg_i}^2\big|_{\sigma^2_{n_{I,i}}=0} = \sigma^2_H \avg{|\gamma_I|^2}_{\hatOmg_i}, \\
    &\mathrm{Kurt}[\Re\sf_{I,i}] := \nonumber\\
    & \int (\Re\sf_{I,i})^4 p(\sf_i|\sigma^2_{n_{I,i}}=0,\sigma^2_H)\dd\sf_{I,i} - 3\Var[\Re\sf_{I,i}]\big|^2_{\sigma^2_{n_{I,i}}=0} \nonumber\\
    &\quad =3\sigma^4_H \big( \avg{|\gamma_{I}|^4}_{\hatOmg_i} - \avg{|\gamma_I|^2}_{\hatOmg_i}^2 \big)
    \equiv 3\sigma^4_H\Gamma_{|\gamma_I|^2|\gamma_I|^2} > 0,
\end{align}
which suggests that the fourth-order distribution property of the signals $\sf_{I,i}$ are determined by the second-order covariance matrix $\Gamma_{ab}$.

Fig.~\ref{fig_sim_compared_dist} provides an intuitive illustration of how the antenna pattern function deforms the distribution of strain amplitude. Assuming the Hanford detector $H_1$ as an example, we simulate a GWB with the original signal strain amplitude drawn from a Gaussian distribution, and examine their projected distributions in the detector at $10\,\mathrm{Hz}$, assuming an isotropic source distribution for $p_\Omega(\hatOmg_i)$. In the figure, we can observe that the original distribution $\Re\hf_+$ is deformed after the projection $\tilde{s}_{H_1}=\Re(\gamma_{H_1}\hf_{H_1})$, corresponding to the gray and green histograms, respectively. We also plot three theoretical probability density functions: the full integral expression of Eq.~\eqref{eq:likelihood_s}, denoted as $\mathcal{G}_I$; the leading-order expansion (the first term of Eq.~\eqref{eq_Taylor_L}), which reduces to the Gaussian distribution $\mathcal{G}$; and the corrected distribution $\mathcal{G}_C$ (the first and second terms of Eq.~\eqref{eq_Taylor_L}). The projected signal exhibits a clear deviation from Gaussianity, showing pronounced positive kurtosis, and inclusion of the second-order correction significantly improves agreement with the simulated distribution.

For stochastic backgrounds, cross-correlation depends on the combined response of two detectors, quantified by the overlap reduction function. This function incorporates the phase difference between spatially separated detectors, which varies with GW frequency. Consequently, the projection effect is inherently frequency dependent. 

In Fig.~\ref{fig_trig_gamma}, assuming an isotropic source distribution, we show the components of the GW detector projection vector, $\vec{y}=(|\gamma_1|^2,|\gamma_2|^2,\Re\gamma_{12},\Im\gamma_{12})$, which directly characterize the projection effect on a stochastic background. The values are computed assuming a uniform source distribution for two detector pairs: LIGO–Livingston vs. LIGO–Hanford, and LIGO–Livingston vs. Virgo. The left and right panels correspond to $10\,\mathrm{Hz}$ and $100\,\mathrm{Hz}$, respectively. All quantities exhibit distinctive, non-trivial structures, and the detector response distribution varies significantly with both the choice of detector pair and the frequency.

Fig.~\ref{fig_corr_compare} illustrates the frequency dependence of the covariance matrix elements. For reference, the sky-averaged overlap reduction function, $\langle \gamma_{IJ}(f, \hatOmg_i) \rangle_{\hatOmg_i}$, is shown in the left panel. We recall that it is related to the usual sky-averaged overlap reduction function by $\langle \gamma_{IJ}(f, \hatOmg_i) \rangle_{\hatOmg_i} = 2 \bar\gamma_{12}(f)/5$. The right panel displays the elements of the second-order covariance matrix $\Gamma_{ab}$. The (3,3) element, corresponding to the variance of $\Re \gamma_{12}$, shows little variation with frequency and eventually stabilizes at a non-zero value. In contrast, (1,3) and (2,3) elements, corresponding to the covariances between $|\gamma_1|^2$ or $|\gamma_2|^2$ and $\Re \gamma_{12}$, exhibit frequency-dependent behavior similar to that of the overlap reduction function. Although not shown in the figure, the fourth component, $\Im(\gamma_{12})$, exhibits a similar trend.

We note that Eq.~(\ref{eq_Taylor_L}) is equivalent to an Edgeworth expansion, which is an asymptotic rather than a convergent series. Thus, including higher-order terms does not necessarily improve the accuracy of the approximation, and the accuracy of the expansion is ultimately determined by the distribution of the antenna pattern function. As shown in Fig.~\ref{fig_trig_gamma}, the antenna-pattern distribution has a non-trivial shape, and the corresponding higher-order cumulants are not guaranteed to decrease with increasing order. Consequently, there is no a priori reason to expect a systematic improvement from the inclusion of higher-order terms. However, as seen in Fig.~\ref{fig_sim_compared_dist}, the second-order correction already provides excellent agreement with the numerical results. For this reason, we retain terms up to second order in the simulations presented in this paper.

%%%%%%%%%%%%%%%%%%%%%%%%%%%%%%%%%%%%%%%%%%%%%%%%%%%%%%%%%%%
%%%%%%%%%%%%%%%%%%%%%%%%%%%%%%%%%%%%%%%%%%%%%%%%%%%%%%%%%%%
\subsection{Effect of anisotropy}\label{subsec_3_2}

%%%%%%%%%%%%%%%%%%%%%%%%%%%%%%%%%%%%%%%%%%%%%%%%%%%%%%%%
\begin{figure}
\begin{tabular}{c}
\includegraphics[width=0.5\textwidth]{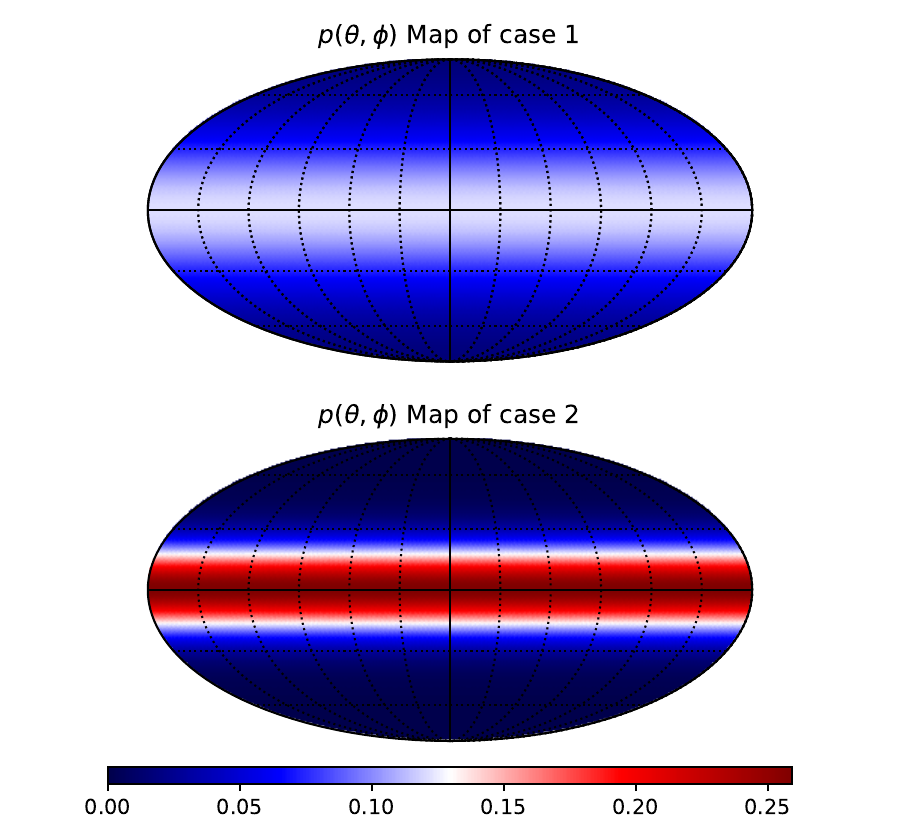}
\end{tabular}
\caption{A sky map illustrating the distribution of the toy-model anisotropic GWB. The top panel corresponds to the $a_1$ model, characterized by a non-uniform distribution with 
$\sigma^2_\Omega = 0.25$, while the bottom panel shows the $a_2$ model with $\sigma^2_\Omega = 0.06$}.
\label{fig_ani_dist}
\end{figure}
%%%%%%%%%%%%%%%%%%%%%%%%%%%%%%%%%%%%%%%%%%%%%%%%%%%%%%%%

%%%%%%%%%%%%%%%%%%%%%%%%%%%%%%%%%%%%%%%%%%%%%%%%%%%%%%%%
\begin{figure*}[t]
\begin{tabular}{cc}
\includegraphics[width=0.5\textwidth]{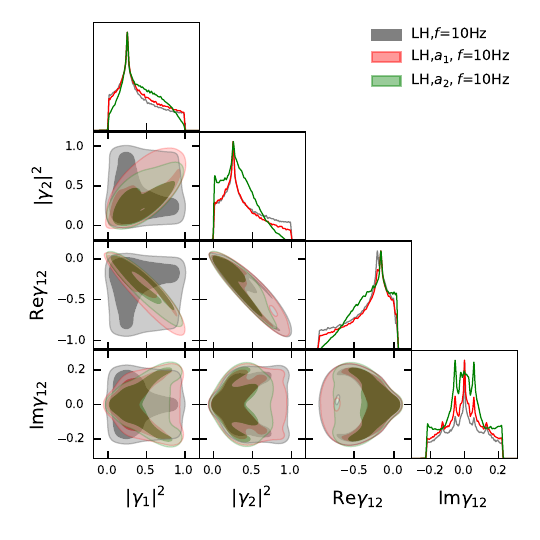}
\includegraphics[width=0.5\textwidth]{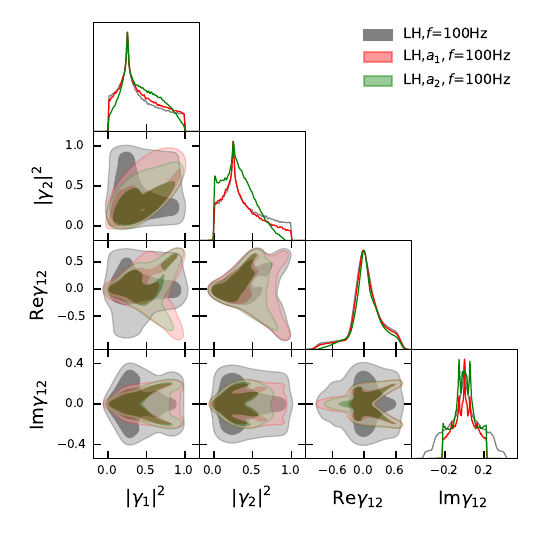}
\end{tabular}
\caption{Distributions of $\vec{y}=(|\gamma_1|^2,|\gamma_2|^2,\Re\gamma_{12},\Im\gamma_{12})$, comparing the cases of isotropic and anisotropic source distributions. Grey indicates the isotropic distribution, red corresponds to the anisotropic distribution case $a_1$ with $\sigma^2_\Omega = 0.25$, and green corresponds to the anisotropic distribution  case $a_2$ with $\sigma^2_\Omega = 0.06$.  The detector pair assumed is LIGO-Livingston vs. LIGO-Hanford. The left and right panels show the results at $10$Hz and $100$Hz, respectively.}
\label{fig_trig_gamma_ani}
\end{figure*}
%%%%%%%%%%%%%%%%%%%%%%%%%%%%%%%%%%%%%%%%%%%%%%%%%%%%%%%%

%%%%%%%%%%%%%%%%%%%%%%%%%%%%%%%%%%%%%%%%%%%%%%%%%%%%%%%%
\begin{figure*}[t]
\begin{tabular}{c}
\includegraphics[width=\textwidth]{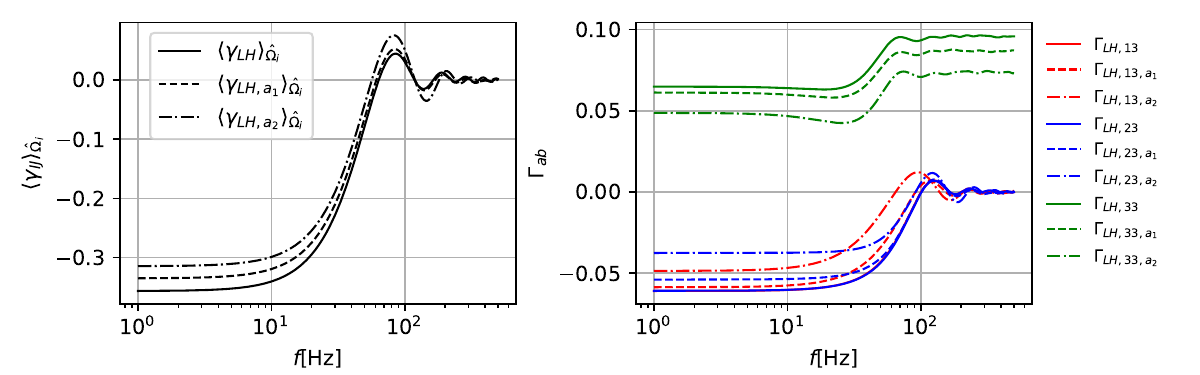}
\end{tabular}
\caption{Frequency dependence of the mean and the second-order covariance matrix elements of $\vec{y}$, comparing isotropic and anisotropic source distribution assumptions. Left panel: The sky-averaged overlap reduction function of $\langle \gamma_{IJ}(f, \hatOmg_i) \rangle_{\hatOmg_i}$. Right panel: The second-order covariance matrix elements, with different colors representing different components. Solid lines correspond to the uniform distribution, dashed lines to the anisotropic distribution case $a_1$ with $\sigma^2_\Omega = 0.25$, and dash-dotted lines to the anisotropic distribution case $a_2$} with $\sigma^2_\Omega = 0.06$.
\label{fig_corr_compare_ani}
\end{figure*}
%%%%%%%%%%%%%%%%%%%%%%%%%%%%%%%%%%%%%%%%%%%%%%%%%%%%%%%%

The choice of the form of $p_\Omega$ is a prior assumption. 
In this section, we evaluate the impact of a non-isotropic $p_\Omega$ on the observable vector $\vec{y}$. While a uniform distribution is a reasonable assumption for extragalactic BBH systems, 
%future scenarios involving observation of white dwarf background with space detectors~\cite{Fryer:2011cx} or exotic models such as clustering of sub-solar mass primordial black hole binaries~\cite{vanDie:2024htf,Afroz:2025urb} or cosmic strings~\cite{Regimbau:2011bm} may necessitate consideration of the highly non-uniform spatial distribution of sources within the Milky Way. 
we may have anisotropic intermittent GW backgrounds. Examples include source populations that follow the highly non-uniform spatial distribution of the Milky Way. Such situation may arise in exotic scenarios such as clustering of sub-solar mass primordial black hole binaries~\cite{vanDie:2024htf,Afroz:2025urb} or cosmic strings~\cite{Chernoff:2009tp,Regimbau:2011bm}. This may also be relevant for the unresolved Galactic double-white-dwarf foreground~\cite{Ungarelli:2001xu,Adams:2013qma} at the high-frequency end of the LISA band and at decihertz frequencies~\cite{Buscicchio:2024wwm,Weaver:2026qgr}, where the source population can become sparse, with only a small number of binaries contributing to each frequency bin. In such a regime, a background formed by sub-threshold events may exhibit intermittent behavior.

To quickly illustrate the effect of anisotropy, we introduce a simple toy model that mimics the structure of the galactic plane:
\begin{align}
p_\Omega \sim \exp\left[-\frac{\cos^2\theta}{2\sigma^2_\Omega}\right],
\end{align}
where $\theta = \pi/2 - \delta$ is the polar angle, which plays the role of the galactic colatitude in this toy model, with the plane lying at $\theta = \pi/2$. Based on current observations, the luminous mass distribution of the Milky Way decreases significantly around $\theta \sim 10^\circ$. 
We consider two distributions with 
$\sigma^2_\Omega = 0.25$ and $0.06$, labeled as $a_1$ and $a_2$, respectively, representing a more diffuse and a more concentrated distribution. These two distributions are shown in Fig.~\ref{fig_ani_dist}.

In Fig.~\ref{fig_trig_gamma_ani}, we present the distribution of $\vec{y}$ for the LIGO-Livingston vs. LIGO-Hanford detector pair under two anisotropic source distributions. For comparison, we also plot results for an isotropic distribution. The figure indicates that anisotropy in the source distribution has only a minor impact on the distribution of $\vec{y}$. Noticeable deviations occur mainly in the tails of the distribution, particularly for the $a_2$ model, which exhibits strong anisotropy. Fig.~\ref{fig_corr_compare_ani} compares the frequency dependence of the overlap reduction function and the second-order covariance matrix elements between the isotropic case and two anisotropic models. The $a_2$ model exhibits up to a 30\% variation in these quantities.

%%%%%%%%%%%%%%%%%%%%%%%%%%%%%%%%%%%%%%%%%%%%%%%%%%%%%%%%%%%
%%%%%%%%%%%%%%%%%%%%%%%%%%%%%%%%%%%%%%%%%%%%%%%%%%%%%%%%%%%
\section{Simulated signal detection and parameter estimation}
\label{sec_4}

Here, we simulate the detection of an intermittent GWB signal using the \texttt{pygwb} module~\cite{Renzini:2023qtj} to demonstrate that neglecting the antenna pattern function can introduce biases in parameter estimation. First, we consider an isotropic source distribution, and then we investigate whether anisotropy in the source distribution influences parameter estimation.

\subsection{Simulation setup}
\label{subsec_4_1}

We consider the antenna pattern functions of LIGO-Livingston and LIGO-Hanford. The data segment length is set to $\Tseg = 4$s, and the total number of data segments is $N_{\rm seg}=10^5$, corresponding to an observation time of approximately four days.  We assume the duty cycle to be $\xi = 0.005$, so the number of segments containing GW signals is $500$. 
This choice is motivated by current estimates of the duty cycle expected for astrophysical backgrounds in ground-based GW detectors. Population studies based on the latest GW catalogs indicate that the astrophysical GWB is likely dominated by binary black hole mergers, with the contribution from binary neutron stars expected to be subdominant~\cite{LIGOScientific:2025bgj}. Correspondingly, several studies predict a duty cycle of order $10^{-3}$--$10^{-2}$ (or smaller depending on the detector sensitivity) for the binary black hole background~\cite{LIGOScientific:2017zlf,Braglia:2022icu,Bagui:2026hrf}.

Following~\cite{Lawrence:2023buo}, we assume that the original distribution of GW signals to be Gaussian and the signal power spectrum is given by $P_H \propto f^{-7/3}$, then the GWB density parameter is given by
\begin{align}
    \Omgw^{\rm int} (f) = \xi\Omref\left(\frac{f}{\fref}\right)^{2/3},
\end{align}
where 
$\Omref$ denotes the population-averaged energy-density amplitude evaluated at the reference frequency $\fref=25\,\Hz$. We generate mock data by injecting Gaussian bursts into randomly selected segments, such that each signal-containing segment is occupied exclusively by a single burst-like signal. The Fourier amplitude of each burst is drawn from a Gaussian distribution with variance $\sigma_H^2(f) \propto P_H(f)$ (see Eq.~\eqref{eq:sigmaH2}). The detector response is then incorporated by applying the antenna pattern functions of the Livingston and Hanford detectors for a randomly chosen sky location.  

A key assumption of our simulation is that there is no correlation between neighboring segments. In reality, if signals are injected at arbitrary times, some may span segment boundaries, although the probability of this occurring is small because we adopt a segment duration of $4$s, which is sufficiently longer than the typical BBH signal duration of $\sim 1$s. Nevertheless, such events may introduce ambiguities in the estimation of the duty cycle and create a degeneracy between $\xi$ and $\sigma_H^2$.

To eliminate this source of uncertainty and focus on the effect of the antenna pattern functions, we require each simulated GW signal to be fully contained within a single data segment. Specifically, we first generate a Gaussian-burst signal occupying one segment in one detector and then construct the corresponding signal in the second detector by applying the appropriate time delay for a given source location. Because the time delay shifts the signal relative to the segment boundaries, a small portion of the signal may extend beyond one edge of the segment. In such cases, the overflowing part is wrapped around and inserted at the opposite edge of the segment. This procedure has a negligible impact on the results, since the maximum time delay between Hanford and Livingston is approximately $10\,\mathrm{ms}$, which is much shorter than the segment duration.

We define the averaged correlated SNR for a single segment $\rho_{\seg}$ as
\begin{align}
\rho_{\seg} := \sqrt{ 2T_{\seg}\sum_{I,J}\int_{f_0}^{\fmax} \frac{\gamma^2_{IJ}(f)P_H^2(f)}{P_{n_I}(f)P_{n_J}(f)}\dd{f}},
\end{align}
where $P^2_H(f)$ is the averaged (squared) spectrum of the GW signal, obtained by averaging the GW signal spectrum over all segments, including both signal-containing and noise-only segments. Here, we choose $f_0 = 10$Hz, and assume identical white noise for both detectors.

Note that our definition of $\rho_{\rm seg}$ differs from that of Ref.~\cite{Lawrence:2023buo}, where the corresponding quantity denotes the expected SNR of a signal-containing segment. To avoid confusion, we denote the latter by $\rho_{\rm burst}$. The two quantities are related by $\rho_{\rm seg}=\xi\rho_{\rm burst}$. 
The total signal-to-noise ratio accumulates over independent segments. Since the mean signal grows linearly with $N_{\rm seg}$ whereas the noise increases only as $\sqrt{N_{\rm seg}}$, the total signal-to-noise ratio scales as 
\begin{align} 
\rho_{\rm tot} \simeq \sqrt{N_{\rm seg}}\,\rho_{\rm seg} = \xi\sqrt{N_{\rm seg}}\,\rho_{\rm burst}. 
\end{align}

In the following simulations, we set $\Omref \approx 4.7\times 10^{-5}$, and the noise level is chosen such that $\rho_{\seg}^2 = (0.1, 0.3, 0.5)$.
These values are chosen to yield total SNRs of approximately $\rho_{\rm tot}\simeq(100, 173, 224)$ for our fiducial setup with $N_{\rm seg}=10^5$. The chosen SNR is relatively high, primarily to improve visual clarity, as larger statistical fluctuations at lower SNR make the underlying trends more difficult to identify. In addition, stronger degeneracies between the parameters tend to slow down the parameter inference. We have nevertheless verified that simulations with lower SNR values yield qualitatively consistent results and do not alter our main conclusions.
%Under this configuration, most of the GW signal intensity lies below the noise level. 
To reduce computation time, we selected only 5 frequency bins between $10$Hz and $100$Hz. 
The validity of this frequency-bin reduction, as well as the assumption of a fixed white-noise spectrum, is examined in Appendix~\ref{app3}.

We generate detector noise that is stationary, Gaussian distributed, and uncorrelated between detectors. While correlated noise sources such as Schumann resonances may become relevant for future detectors, they are persistent rather than intermittent and are therefore expected to mainly reduce the signal-to-noise ratio rather than bias an intermittent-signal search. Their impact is thus expected to be small.

%%%%%%%%%%%%%%%%%%%%%%%%%%%%%%%%%%%%%%%%%%%%%%%%%%%%%%%%
\begin{figure*}
\begin{tabular}{cc}
\includegraphics[width=0.5\textwidth]{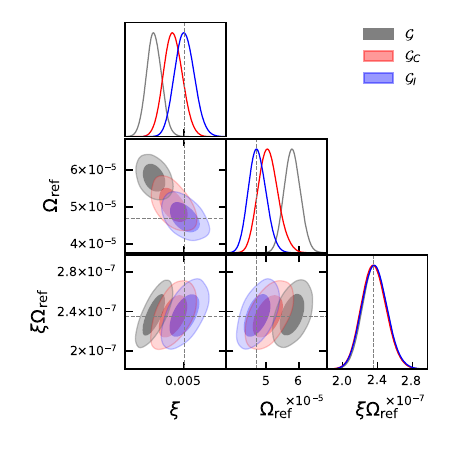}
\includegraphics[width=0.5\textwidth]{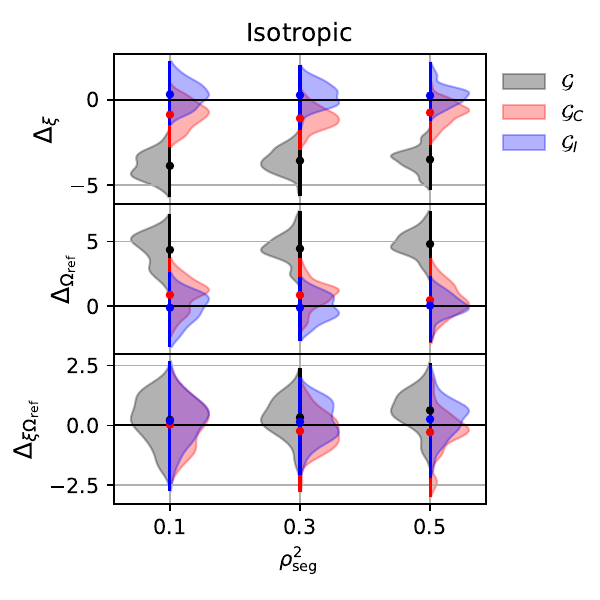}
\end{tabular}
\caption{ Result of the isotropic case. 
Left panel: Posterior probability distribution of the parameters obtained from a single simulation with $\rho^2_{\seg}=0.1$.
Right panel: Comparison of the biases $(\Delta_\xi,\Delta_{\Omref},\Delta_{\xi\Omref})$ for three likelihood models: the uncorrected Gaussian model ($\Gaus$, black), the model with numerically integrated $p_\Omega$ ($\Gaus_I$, blue), and the model with covariance matrix $\Gamma_{ab}$ corrections ($\Gaus_C$, red). The histograms are shown for different SNRs $\rho_{\seg}^2 \sim (0.1,0.3,0.5)$, each based on results from 50 simulations.
}
\label{fig_histogram_bias_isotropic}
\end{figure*}
%%%%%%%%%%%%%%%%%%%%%%%%%%%%%%%%%%%%%%%%%%%%%%%%%%%%%%%%

We consider three models for comparison: the original Gaussian likelihood model (labeled $\Gaus$), the likelihood model with a full integral over $\hatOmg$, as given in Eq.~\eqref{eq:cov} (labeled $\Gaus_I$), and the likelihood model expanded to include the second-order covariance matrix, as given in Eq.~\eqref{eq_Taylor_L} (labeled $\Gaus_C$). Since the $\Gaus_I$ model involves integration, sampling process is more computationally demanding. Therefore, for this comparison, we exclude the estimation of the noise parameters for the two detectors, assuming their noise levels are known, and focus solely on parameter estimation of the two parameters: the duty cycle $\xi$ and the signal amplitude $\Omref\propto\sigma^2_H$. We assume uniform prior for $\xi$ and log-uniform prior for $\Omref$,
\begin{align}
    & \xi \sim U(0,1),\ \log_{10}\Omref\sim U(-10, 0).
\end{align}

To assess the impact of stochastic variations in the simulations, we perform 50 repeated simulations. For each posterior distribution obtained from a single run, we compute the biases for each parameter $(\Delta_\xi,\Delta_{\Omref})$ as well as the bias of the total GWB density $\Delta_{\xi\Omref}$. Specifically, the bias $\Delta_X$ of parameter $X$ is defined as 
\begin{align}
    \Delta_X := \frac{\avg{X}_{\mathrm{post}}-X_{\mathrm{inj}}}{\sigma_{X,\mathrm{post}}},
\end{align}
where $\avg{X}_{\mathrm{post}}$ is the posterior mean of $X$, $X_{\mathrm{inj}}$ is the injected parameter value, and $\sigma^2_{X,\mathrm{post}}$ is the variance of the marginalized posterior distribution of $X$. This quantity measures how many standard deviations ($\sigma$) the inferred parameter value deviates from the true value, thereby characterizing the intrinsic capability of each model to recover the true parameters. 
A value of $|\Delta_X| < 1$ indicates that the residual bias is smaller than the expected statistical uncertainty and is therefore subdominant to statistical errors for the experimental sensitivities considered.

\subsection{Isotropic source distribution case}
\label{subsec_4_2}

We first consider the case in which GW sources are isotropically distributed over the sky. Specifically, we draw the source coordinates and polarization angle from
\begin{align}
    \alpha\sim U(0,2\pi),\ \cos{\left(\frac{\pi}{2}-\delta\right)}\sim U(-1,1),\ \psi\sim U(0,\pi).\nonumber
\end{align}

In Fig.~\ref{fig_histogram_bias_isotropic}, the left panel shows representative posterior distributions from a single selected simulation, while the right panel presents the bias distributions for each parameter $(\Delta_\xi,\Delta_{\Omref},\Delta_{\xi\Omref})$ obtained from 50 simulations for different SNRs $\rho_{\seg}^2 \sim (0.1,0.3,0.5)$. Results from multiple simulations on the right panel clearly indicate that the uncorrected Gaussian model $\mathcal{G}$ introduces systematic biases regardless of SNR: specifically, $\xi$ tends to be underestimated, while $\Omref$ is overestimated. The model using the full integral form, $\mathcal{G}_I$, effectively corrects these biases. We also find that the model incorporating the second-order covariance matrix correction, $\mathcal{G}_C$, produces results that are closer to those of the full integral model. 

Another notable observation is that the overall posterior variance of the total GWB energy density, $\Omgw^{\rm int}\propto\xi\Omref$, remains essentially unbiased across all models. This is also seen in the left panel of Fig.~\ref{fig_histogram_bias_isotropic}, where the degeneracy between $\xi$ and $\Omref$ appears to align along a direction that preserves the same $\Omgw^{\rm int}$ amplitude, specifically satisfying the relation $\xi \Omref = \text{const}$. 

%%%%%%%%%%%%%%%%%%%%%%%%%%%%%%%%%%%%%%%%%%%%%%%%%%%%%%%%
\begin{figure}
\begin{tabular}{c}
\includegraphics[width=0.5\textwidth]{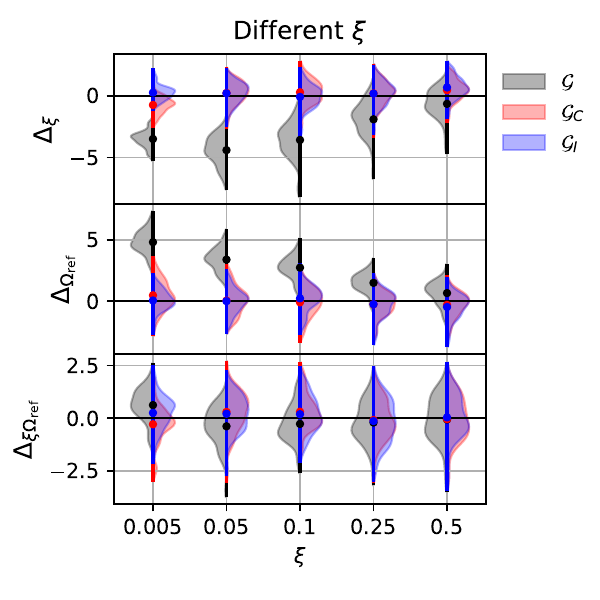}
\end{tabular}
\caption{Similar to the right panel of Fig.~\ref{fig_histogram_bias_isotropic}, the histogram of the biases is shown for different values of $\xi$. The SNR is fixed at $\rho^2_{\rm seg}=0.5$. }
\label{fig_histogram_bias_isotropic_differentxi}
\end{figure}
%%%%%%%%%%%%%%%%%%%%%%%%%%%%%%%%%%%%%%%%%%%%%%%%%%%%%%%%

%%%%%%%%%%%%%%%%%%%%%%%%%%%%%%%%%%%%%%%%%%%%%%%%%%%%%%%%
\begin{figure*}
\begin{tabular}{cc}
\includegraphics[width=0.5\textwidth]{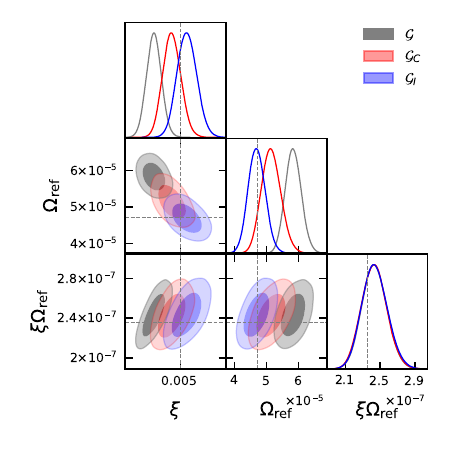}
\includegraphics[width=0.5\textwidth]{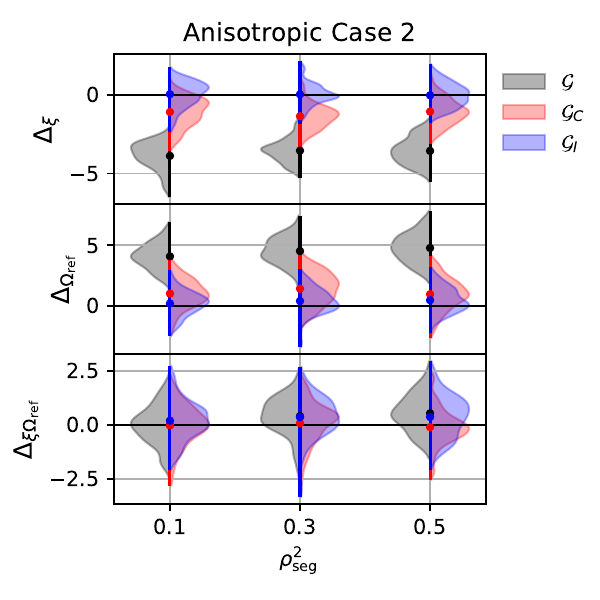}
\end{tabular}
\caption{Same as Fig.~\ref{fig_histogram_bias_isotropic}, except the data are generated under an anisotropic background assumption, while the analysis assumes an isotropic background.}
\label{fig_histogram_bias_anisotropic}
\end{figure*}
%%%%%%%%%%%%%%%%%%%%%%%%%%%%%%%%%%%%%%%%%%%%%%%%%%%%%%%%

We next perform tests for different values of the duty cycle. Specifically, we consider five cases, $\xi=0.005$, $0.05$, $0.1$, $0.25$, and $0.5$, while keeping the total number of segments containing GW signals fixed to be the same as in the $\xi=0.005$ case. At the same time, we adjust the noise levels such that $\rho^2_{\seg}=0.5$ remains unchanged across all cases. The results are shown in Fig.~\ref{fig_histogram_bias_isotropic_differentxi}. We find that the bias in the uncorrected Gaussian model $\Gaus$ decreases as $\xi$ increases, both for $\xi$ and $\Omega_{\rm ref}$. This is because a larger $\xi$ corresponds to weaker non-Gaussianity in the signal, thereby reducing the relative advantage of the full-likelihood formulation. From our simulations, we find that the uncorrected Gaussian model becomes effectively unbiased for $\xi \sim 0.5$.

We note that the improved intermittent-search models proposed here, $\Gaus_C$ and $\Gaus_I$, are both based on the assumption that the angular dependence of the GW power spectrum can be factorized as
$\Hf_{P,i}(f,\hatOmg)=\Hf_P(f)\delta^2(\hatOmg,\hatOmg_i)$.
However, in the high-duty-cycle regime, the probability that multiple GW signals overlap within the same data segment increases. In such cases, the factorization assumption is no longer strictly valid, which may explain the increasing bias observed at higher duty cycles. In this regime, the standard cross-correlation search currently employed by the LVK~\cite{LIGOScientific:2025bgj} is expected to achieve comparable performance~\cite{Drasco:2002yd,Lawrence:2023buo}.

\subsection{Anisotropic source distribution}
\label{subsec_4_3}

A natural question that arises is whether anisotropy in the GWB introduces additional bias in parameter estimation. To investigate this, we perform simulations using mock data generated under an anisotropic GWB assumption, while analyzing the data under the simplifying assumption of isotropy.

First, we create the data assuming the anisotropic distribution equivalent to the $a_2$ model described in Sec.~\ref{subsec_3_2},
\begin{align}
    \alpha\sim U(0,2\pi),\ \cos{\left(\frac{\pi}{2}-\delta\right)}\sim \Gauss[\sigma^2_\Omega]_{(-1,1)},\ \psi\sim U(0,\pi),\nonumber
\end{align}
where $\Gauss[\sigma^2_\Omega]_{(-1,1)}$ denotes the Gaussian distribution truncated within $(-1,1)$, and $\sigma^2_\Omega=0.06$. 

As in the previous subsection, we perform detection simulations for three different SNR cases, $\rho^2_\seg = (0.1, 0.3, 0.5)$, repeating each scenario 50 times. The resulting bias in the posterior probability distribution is shown in Fig.~\ref{fig_histogram_bias_anisotropic}. This result shows almost no difference compared to the isotropic case shown in Fig.~\ref{fig_histogram_bias_isotropic}, indicating that even a strong level of anisotropy has no significant impact on parameter estimation. Therefore, in terms of parameter estimation accuracy, assuming an isotropic prior $p_\Omega = 1/(4\pi)$ is sufficient.

\begin{figure}
\begin{tabular}{c}
\includegraphics[width=0.5\textwidth]{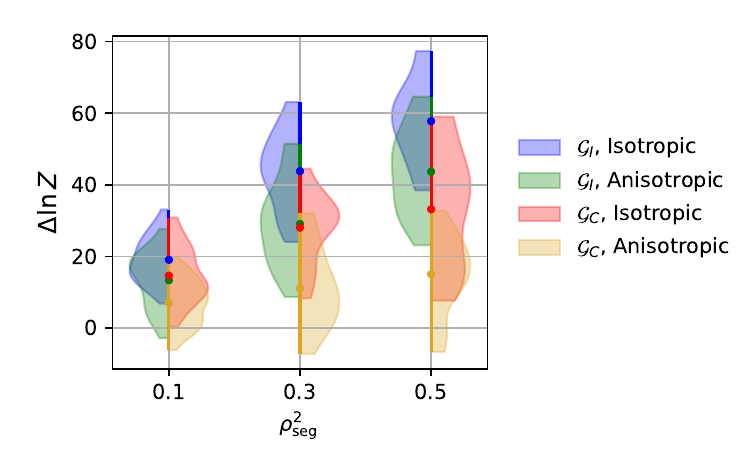}
\end{tabular}
\caption{Histograms of the log Bayes factors, $\ln{Z_{\mathcal{G}_I}/Z_{\mathcal{G}}}$ and $\ln{Z_{\mathcal{G}_C}/Z_{\mathcal{G}}}$, are shown for isotropic and anisotropic cases for different SNRs. }
\label{fig_histogram_lnZ}
\end{figure}

However, we find that differences between isotropic and anisotropic assumptions appear in the Bayes factors. In Fig.~\ref{fig_histogram_lnZ}, we compare the Bayes factors $\Delta\ln Z$ for the $\mathcal{G}_I$ and $\mathcal{G}_C$ models relative to the $\mathcal{G}$ model, based on distributions obtained from 50 simulations each. 
%The Bayes factor of a model $X$ is given by marginalizing the posterior probability distribution $p_{\mathrm{post}}(\pmb{\theta})$ of the model parameters $\pmb\theta$,
%\begin{align}
%    Z_X = \int p_{\mathrm{post}}(\pmb{\theta}) \pi(\pmb\theta)\dd\pmb\theta,
%\end{align}
%where $\pi(\pmb\theta)$ denotes the prior. 
% The ratio $\ln(Z_1/Z_2)$ quantifies the relative detection capability of two models; a higher Bayes factor indicates a more reliable model. 

For a model $X$, the evidence $Z_X$ is obtained by
marginalizing the likelihood over the prior of the model parameters
$\pmb{\theta}$,
\begin{equation}
  Z_X = \int \mathcal{L}^{X}_{\rm tot}(\pmb{s}|\pmb{\theta})\,
             \pi(\pmb{\theta})\,\mathrm{d}\pmb{\theta} ,
  \label{eq:evidence}
\end{equation}
where $\mathcal{L}^{X}_{\rm tot}$ denotes the total likelihood evaluated with the signal likelihood of model $X$,
and $\pi(\pmb{\theta})$ is the prior. The Bayes factor between two models is the evidence ratio $Z_1/Z_2$, and $\ln(Z_1/Z_2)>0$ indicates that the data favor model 1 over model 2. From Fig.~\ref{fig_histogram_lnZ}, we observe that in the isotropic case both $\mathcal{G}_I$ and $\mathcal{G}_C$ are favored over $\mathcal{G}$. Under anisotropic conditions, however, these log Bayes factors are reduced. This suggests that when sources are anisotropically distributed, using isotropic priors does not introduce bias in the parameter estimates; however, it reduces the preference for $\mathcal{G}_I$ and $\mathcal{G}_C$ models over $\mathcal{G}$.

%%%%%%%%%%%%%%%%%%%%%%%%%%%%%%%%%%%%%%%%%%%%%%%%%%%%%%%%
%%%%%%%%%%%%%%%%%%%%%%%%%%%%%%%%%%%%%%%%%%%%%%%%%%%%%%%%

\section{Summary}
\label{sec_5}
The astrophysical GWB exhibits a highly non-Gaussian nature in the frequency band of ground-based detectors, posing unique challenges for analysis. In this work, we highlighted a critical aspect that must be considered when searching for such a non-Gaussian stochastic background: the detector antenna patterns, which enter the analysis through the overlap reduction function in the cross-correlation method.

We pointed out that the non-spherical nature of the antenna pattern functions of detectors modifies the statistical properties of the signal distribution. This effect is frequency-dependent when the two detectors are spatially separated, due to the phase difference between them. We showed that this modification generally introduces positive kurtosis in the signal distribution.

When constructing the likelihood function, this projection effect must be incorporated. However, incorporating it requires an additional integration over the source direction, which significantly increases computational cost for parameter estimation because the likelihood must be evaluated at every sample point. To overcome this, we introduced a method that treats the effect as a higher-order moment expansion by utilizing the covariance matrix of the detector response function.

To compare parameter estimation performance, we simulated the detection and parameter estimation by generating mock data containing Gaussian burst signals. Three likelihood models were tested: the conventional Gaussian model, which averages the antenna pattern effect over the sky; the full integral formulation; and the corrected model with the second-order covariance matrix. The results show that the Gaussian model consistently underestimates the duty cycle and overestimates the variance of the GW amplitude, whereas the full integral formulation exhibits no deviation. We also demonstrated that the second-order correction effectively mitigates the bias in parameter estimation. 

In addition, we investigated the effect of anisotropic source distributions. The detector response function and its covariance matrix show slight differences when anisotropy is present. Finally, we examined whether assuming an isotropic source distribution in the analysis introduces bias in the estimated parameters when the true background is anisotropic. Through detailed simulations, we found no such bias in parameter estimation. However, we observed that the Bayes factors decrease when the source distribution is anisotropic, even though no parameter bias occurs.

This study focused on the impact of detector response in intermittent searches, using simulations based on Gaussian-distributed, burst-like data. In contrast, the most probable astrophysical background signal originates from BBH mergers, which pose additional challenges due to their complex, non-Gaussian characteristics arising from source population characteristics and waveform structures. In future work, we will address more realistic scenarios by modeling CBC sources and systematically tackling these challenges to ensure that we are prepared to analyze the forthcoming observational data.

\section*{Acknowledgments}
We thank David Alonso, Neil Cornish, Federico De Lillo, Jessica Lawrence, Guo Chin Liu, Nick Van Remortel, Arianna Renzini, Joseph Romano, and Takahiro Yamamoto for very helpful discussion. XL and SK are supported by the I+D grant PID2023-149018NB-C42 and the Grant IFT Centro de Excelencia Severo Ochoa No CEX2020-001007-S, funded by MCIN/AEI/10.13039/501100011033, the Leonardo Grant for Scientific Research and Cultural Creation 2024 from the BBVA Foundation, and Japan Society for JSPS KAKENHI Grant no. JP23H00110 and JP24K00624.

\begin{appendix}

\section{Derivation of Eq.~(\ref{eq:likelihood_s})}
\label{app:deriv_likelihood_s}

Here we show how to obtain Eq.~(\ref{eq:likelihood_s}) from Eq.~(\ref{eq:likelihood_s_int}). The derivation proceeds in two steps: (i) the integral over the GW amplitudes is written as a Gaussian integral over the two complex polarization amplitudes and performed analytically; and (ii) the Woodbury matrix identity and the matrix determinant lemma are used to identify the covariance matrix $\mathbb{C}$ of Eq.~(\ref{eq:cov}).

\subsection*{(i) Gaussian integral over the GW amplitudes}

One can rewrite the integral in Eq.~(\ref{eq:likelihood_s_int}) as
\begin{align}
    &\Like_s(\tilde{\pmb{s}}_i) = \prod_k\int p_\Omega(\hatOmg_i)\dd\hatOmg_i \int \prod_I^{N_d} p_{n_I}\bigg( \sf_{I,i,k} \nonumber\\
    &\qquad - \sum_P \tilde{F}_{I,P,i,k}\Hf_{P,k} \bigg) \prod_P p_{\gw}\big( \Hf_{P,k} \big) \dd\Hf_{P,k} \nonumber\\
    &\qquad := \prod_k\int p_\Omega(\hatOmg_i)\Like_{\mathrm{inner}}(\tilde{\pmb{s}},f_k,\hatOmg_i)\dd\hatOmg_i
    \label{eq:Ls}
\end{align}
where we use
\begin{align}
    & \sf_{I,i,k}:=\sf_{I,i}(f_k), \ \Hf_{P,k}:=\Hf_{P}(f_k) \nonumber\\
    & \tilde{F}_{I,P,i,k}:=F_{I,P}(\hatOmg_i)\ee^{-\ii2\pi f_k\hatOmg_i\cdot\pmb{x}_I}
\end{align}
for brevity. For later use, we also define 
\begin{equation}
\sigma_{n_I,k}^2:=\sigma_{n_I}^2(f_k), ~~~ \sigma_{H,k}^2:=\sigma_H^2(f_k)
\end{equation}
Throughout this appendix $I=1,\dots,N_d$ labels the detectors, $P=+,\times$ the two polarization states, $i$ the data segment and $k$ the frequency bin.

The inner integral over $\Hf_{P,k}$ can be evaluated analytically when $\Hf_{P,k}$ follows a Gaussian distribution.  We therefore substitute the following Gaussian distributions
\begin{align}
    & p_{n_I}(n|\sigma^2_n) := \frac{1}{2\pi\sigma^2_n}\exp\bigg(- \frac{|n|^2}{2\sigma^2_n} \bigg), \label{eq:pn}\\
    & p_{\gw}(h|\sigma^2_H) := \frac{1}{2\pi\sigma^2_H}\exp\bigg(- \frac{|h|^2}{2\sigma^2_H} \bigg)\,,
  \label{eq:pgw}
\end{align}
which are normalized with respect to the real and imaginary parts of their complex arguments.

It is convenient to arrange the data, the GW amplitudes and the antenna pattern functions into
\begin{align}
  \tilde{\pmb{s}} &:=
    \left[\tilde{s}_{1,i,k},\dots,\tilde{s}_{N_d,i,k}\right]^T
  \label{eq:Asvec}\\
  \tilde{\pmb{h}} &:=
    \left[\tilde{H}_{+,k},\,\tilde{H}_{\times,k}\right]^T
  \label{eq:Ahvec}
\end{align}
and
\begin{equation}
  \tilde{\mathbf{F}} :=
  \begin{bmatrix}
    \tilde{F}_{1,+,i,k} & \tilde{F}_{1,\times,i,k}\\
    \vdots & \vdots \\
    \tilde{F}_{N_d,+,i,k} & \tilde{F}_{N_d,\times,i,k}
  \end{bmatrix} ,
  \label{eq:AF}
\end{equation}
\begin{equation}
  \mathbf{N} :=
  \begin{bmatrix}
    \sigma_{n_1,k}^2 & & \\
     & \ddots & \\
     & & \sigma_{n_{N_d},k}^2
  \end{bmatrix} ,
  \label{eq:AN}
\end{equation}
where $\tilde{\mathbf{F}}$ is an $N_d\times2$ matrix and $\mathbf{N}$ is an $N_d\times N_d$ diagonal matrix.  The signal projected onto the detectors is then simply $\tilde{\mathbf{F}}\tilde{\pmb{h}}$, and the integration measure is
\begin{equation}
  \prod_P\mathrm{d}\tilde{H}_{P,k} =: \mathrm{d}^4\tilde{\pmb{h}}
  = \prod_P \mathrm{d}\,\mathrm{Re}\,\tilde{H}_{P,k}\,
    \mathrm{d}\,\mathrm{Im}\,\tilde{H}_{P,k} ,
  \label{eq:Ameasure}
\end{equation}
consistently with the normalization of Eqs.~(\ref{eq:pn}) and (\ref{eq:pgw}).

With these definitions the exponent of the integrand of Eq.~(\ref{eq:Ls}) becomes
\begin{align}
  \sum_I^{N_d}&\frac{\big|\tilde{s}_{I,i,k}
    -(\tilde{\mathbf{F}}\tilde{\pmb{h}})_I\big|^2}
    {2\sigma_{n_I,k}^2}
  + \sum_P\frac{\big|\tilde{H}_{P,k}\big|^2}{2\sigma_{H,k}^2}
  \nonumber\\
  &= \frac{1}{2}\left(\tilde{\pmb{s}}
     - \tilde{\mathbf{F}}\tilde{\pmb{h}}\right)^\dagger
     \mathbf{N}^{-1}\left(\tilde{\pmb{s}}
     - \tilde{\mathbf{F}}\tilde{\pmb{h}}\right)
   + \frac{\tilde{\pmb{h}}^\dagger\tilde{\pmb{h}}}
     {2\sigma_{H,k}^2}
  \nonumber\\
  &= \frac{1}{2}\tilde{\pmb{h}}^\dagger\tilde{\mathbf{A}}
     \tilde{\pmb{h}}
   + \mathrm{Re}\big[\tilde{\pmb{b}}^\dagger
     \tilde{\pmb{h}}\big]
   + \frac{1}{2}c ,
  \label{eq:Aexponent}
\end{align}
where the two cross terms have been combined using $\tilde{\pmb{s}}^\dagger\mathbf{N}^{-1}\tilde{\mathbf{F}}\tilde{\pmb{h}}+\tilde{\pmb{h}}^\dagger\tilde{\mathbf{F}}^\dagger\mathbf{N}^{-1}\tilde{\pmb{s}}=2\,\mathrm{Re}[\tilde{\pmb{s}}^\dagger\mathbf{N}^{-1}\tilde{\mathbf{F}}\tilde{\pmb{h}}]$, and we have defined the $2\times2$ Hermitian matrix $\tilde{\mathbf{A}}$, the complex two-component vector $\tilde{\pmb{b}}$ and the scalar $c$ as
\begin{align}
  \tilde{\mathbf{A}} &:= \frac{1}{\sigma_{H,k}^2}\mathbf{I}_{2\times2}
    + \tilde{\mathbf{F}}^\dagger\mathbf{N}^{-1}\tilde{\mathbf{F}} ,
  \label{eq:AAtilde}\\
  \tilde{\pmb{b}} &:=
    -\tilde{\mathbf{F}}^\dagger\mathbf{N}^{-1}\tilde{\pmb{s}} ,
  \label{eq:Abtilde}\\
  c &:= \tilde{\pmb{s}}^\dagger\mathbf{N}^{-1}
    \tilde{\pmb{s}}
    = \sum_I^{N_d}\frac{\left|\tilde{s}_{I,i,k}\right|^2}
    {\sigma_{n_I,k}^2} .
  \label{eq:Ac}
\end{align}
Collecting also the normalization factors of the $N_d$ noise distributions and of the two GW distributions, the inner integral reads
\begin{align}
  &\mathcal{L}_{\rm inner}(\tilde{\pmb{s}},f_k,\hatOmg_i)
  = \frac{1}{(2\pi)^{N_d+2}\sigma_{H,k}^4
     \prod_I^{N_d}\sigma_{n_I,k}^2} \nonumber\\
  &\qquad \times\int\exp\Big(
     -\frac{1}{2}\tilde{\pmb{h}}^\dagger\tilde{\mathbf{A}}
     \tilde{\pmb{h}}
     - \mathrm{Re}\big[\tilde{\pmb{b}}^\dagger
       \tilde{\pmb{h}}\big]
     - \frac{1}{2}c\Big)\,\mathrm{d}^4\tilde{\pmb{h}} .
  \label{eq:app_inner_likelihood}
\end{align}

Since $\tilde{\mathbf{A}}$ is Hermitian and positive definite, the integral can be evaluated analytically via the standard complex Gaussian integration formula
\begin{align}
  \int &\exp\left(-\frac{1}{2}\tilde{\pmb{h}}^\dagger
  \tilde{\mathbf{A}}\tilde{\pmb{h}}
  - \mathrm{Re}\big[\tilde{\pmb{b}}^\dagger
    \tilde{\pmb{h}}\big]
  - \frac{1}{2}c\right)\mathrm{d}^4\tilde{\pmb{h}}
  \nonumber\\
  &= \frac{(2\pi)^2}{\det\tilde{\mathbf{A}}}
  \exp\left[\frac{1}{2}\tilde{\pmb{b}}^\dagger
  \tilde{\mathbf{A}}^{-1}\tilde{\pmb{b}}
  - \frac{1}{2}c\right] .
  \label{eq:exp}
\end{align}
This is obtained by shifting the integration variable to $\tilde{\pmb{u}}=\tilde{\pmb{h}}+\tilde{\mathbf{A}}^{-1}\tilde{\pmb{b}}$, which removes the term linear in $\tilde{\pmb{h}}$ and leaves $\tfrac{1}{2}\tilde{\pmb{b}}^\dagger\tilde{\mathbf{A}}^{-1} \tilde{\pmb{b}}$.

\subsection*{(ii) Identification of the covariance matrix}

Inserting Eqs.~(\ref{eq:Abtilde}) and (\ref{eq:Ac}) into the exponent of
Eq.~(\ref{eq:exp}) gives
\begin{equation}
  \tilde{\pmb{b}}^\dagger\tilde{\mathbf{A}}^{-1}
  \tilde{\pmb{b}} - c
  = -\tilde{\pmb{s}}^\dagger\left(\mathbf{N}^{-1}
  - \mathbf{N}^{-1}\tilde{\mathbf{F}}\tilde{\mathbf{A}}^{-1}
  \tilde{\mathbf{F}}^\dagger\mathbf{N}^{-1}\right)\tilde{\pmb{s}} .
  \label{eq:bAb}
\end{equation}
The bracket in Eq.~(\ref{eq:bAb}) has precisely the structure of the
right-hand side of the Woodbury matrix identity
\begin{align}
  &\left(M + UXV\right)^{-1} \nonumber\\
  &\qquad = M^{-1} - M^{-1}U\left(X^{-1}
    + VM^{-1}U\right)^{-1}VM^{-1} ,
\end{align}
Identifying $M=\mathbf{N}$, $U=\tilde{\mathbf{F}}$ and
$V=\tilde{\mathbf{F}}^\dagger$, the remaining factor is fixed by
\begin{equation}
  X^{-1} + \tilde{\mathbf{F}}^\dagger\mathbf{N}^{-1}\tilde{\mathbf{F}}
  = \tilde{\mathbf{A}}
  = \frac{1}{\sigma_{H,k}^2}\mathbf{I}_{2\times2}
  + \tilde{\mathbf{F}}^\dagger\mathbf{N}^{-1}\tilde{\mathbf{F}} ,
\end{equation}
that is, $X=\sigma_{H,k}^2\mathbf{I}_{2\times2}$.  The bracket in
Eq.~(\ref{eq:bAb}) is therefore the inverse of the $N_d\times N_d$ matrix
\begin{equation}
  \mathbb{C} := M + UXV
  = \mathbf{N} + \sigma_{H,k}^2
  \tilde{\mathbf{F}}\tilde{\mathbf{F}}^\dagger ,
    \label{eq:ACov}
\end{equation}
which is exactly the covariance matrix in Eq.~(\ref{eq:cov}). The
exponential term thus takes the standard form
\begin{equation}
  \tilde{\pmb{b}}^\dagger\tilde{\mathbf{A}}^{-1}
  \tilde{\pmb{b}} - c
  = -\tilde{\pmb{s}}^\dagger\mathbb{C}^{-1}\tilde{\pmb{s}} .
  \label{eq:sCs}
\end{equation}

The determinant of $\tilde{\mathbf{A}}$ is obtained from the same
decomposition via the matrix determinant lemma
\begin{equation}
  \det\left(Y + UV\right)
  = \det Y \,\det\left(I + VY^{-1}U\right) \,.
  \label{eq:A27}
\end{equation}
Applying it to Eq.~(\ref{eq:ACov}) with
$U=\sigma_{H,k}^2\tilde{\mathbf{F}}$ and
$V=\tilde{\mathbf{F}}^\dagger$, and using
$\mathbf{I}_{2\times2}+\sigma_{H,k}^2\tilde{\mathbf{F}}^\dagger
\mathbf{N}^{-1}\tilde{\mathbf{F}} = \sigma_{H,k}^2\tilde{\mathbf{A}}$
together with $\det(\sigma_{H,k}^2\tilde{\mathbf{A}})
= \sigma_{H,k}^4\det\tilde{\mathbf{A}}$, which holds because
$\tilde{\mathbf{A}}$ is a $2\times2$ matrix, one finds
\begin{align}
  \det\mathbb{C}
  &= \det\mathbf{N}\,\det\left(\mathbf{I}_{2\times2}
     + \sigma_{H,k}^2\tilde{\mathbf{F}}^\dagger\mathbf{N}^{-1}
     \tilde{\mathbf{F}}\right) \nonumber\\
  &= \sigma_{H,k}^4\left(\prod_I^{N_d}\sigma_{n_I,k}^2\right)
     \det\tilde{\mathbf{A}} ,
  \label{eq:A28}
\end{align}
that is,
\begin{equation}
  \frac{1}{\det\tilde{\mathbf{A}}}
  = \sigma_{H,k}^4\left(\prod_I^{N_d}\sigma_{n_I,k}^2\right)
    \frac{1}{\det\mathbb{C}} .
  \label{eq:detA}
\end{equation}

Substituting Eqs.~(\ref{eq:exp}), (\ref{eq:sCs}) and (\ref{eq:detA}) into
the inner likelihood~(\ref{eq:app_inner_likelihood}), all the noise and signal variances
cancel against the prefactor,
\begin{align}
  \frac{1}{(2\pi)^{N_d+2}\sigma_{H,k}^4\prod_I\sigma_{n_I,k}^2}
  &\cdot\frac{(2\pi)^2}{\det\tilde{\mathbf{A}}}
  = \frac{1}{(2\pi)^{N_d}\det\mathbb{C}} \nonumber\\
  &= \frac{1}{\det\left[2\pi\mathbb{C}\right]} ,
\end{align}
where the last equality follows because $\mathbb{C}$ is an
$N_d\times N_d$ matrix.  One therefore obtains
\begin{align}
  \mathcal{L}_s(\tilde{\pmb{s}}_i)
  = \prod_k \int
  &\frac{p_\Omega(\hatOmg_i)\mathrm{d}\hatOmg_i}
  {\det\left[2\pi\mathbb{C}(f_k,\hatOmg_i)\right]} \nonumber\\
  &\times\exp\left(-\frac{1}{2}\tilde{\pmb{s}}^\dagger(f_k)
  \mathbb{C}^{-1}(f_k,\hatOmg_i)
  \tilde{\pmb{s}}(f_k)\right) ,
\end{align}
which is Eq.~(\ref{eq:likelihood_s}).

\section{Validation of our reduced setup}\label{app3}

%%%%%%%%%%%%%%%%%%%%%%%%%%%%%%%%%%%%%%%%%%%%%%%%%%%%%%%%
\begin{figure}[t]
\begin{tabular}{c}
\includegraphics[width=0.5\textwidth]{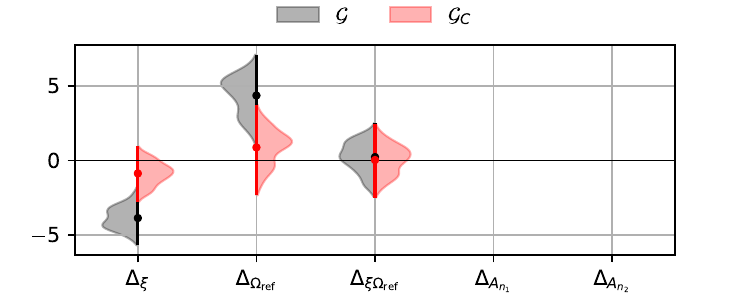}\\
\includegraphics[width=0.5\textwidth]{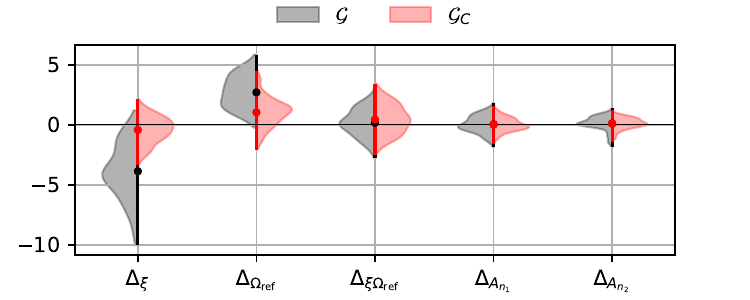}
\end{tabular}
\caption{ 
Distribution of the biases from 50 simulations. The SNR and duty cycle are $\rho^2_{\seg}=0.1$ and $\xi=0.005$, respectively. The top panel is the same as that shown in Fig.~\ref{fig_histogram_bias_isotropic}, while the bottom panel shows the corresponding results obtained using the full set of frequency bins in the range $f\in[10,100]$ and the realistic Advanced-LIGO noise PSD $P_0(f)$. In the latter case, the Bayesian inference includes two additional free parameters, namely the overall noise-amplitude parameters for each detector, $A_{n_I}$, defined through $P_{n_I}(f)=A_{n_I}P_0(f)$.
}
\label{fig_histogram_bias_with_frequency}
\end{figure}
%%%%%%%%%%%%%%%%%%%%%%%%%%%%%%%%%%%%%%%%%%%%%%%%%%%%%%%%

In our simulations, to make exploring a large number of realizations computationally feasible, we adopt a simplified setup consisting of white noise and a reduced number of frequency bins. As a result, potential effects associated with the frequency dependence of the overlap reduction function and its corrections may be underestimated. In addition, we assume that the noise level is perfectly known and do not perform a joint inference of the signal and noise parameters. This approximation may break down when some segments are signal dominated, such cases may require simultaneous fitting of the noise spectrum and signal parameters particularly in the presence of non-stationary noise. In this appendix, we assess the validity of these simplifications and examine whether they can affect our main conclusions.

We now repeat the simulations using realistic noise PSD and a larger number of frequency bins and find that the overall behavior is essentially unchanged from the simplified case. We use all frequency bins within the full frequency range $f\in [10, 100]$Hz at the chosen sampling rate, together with the colored advanced-LIGO PSD $P_0(f)$. The noise is modeled using a single amplitude parameter $A_{n_I}$, such that $P_{n_I}(f)=A_{n_I}P_0(f)$. This remains a simplified setup, as we assume that the shape of the noise spectrum is known and allow only its overall amplitude to vary. Nevertheless, it provides a useful test of the impact of noise estimation. The segment SNR is fixed to $\rho^2_{\seg}=0.1$, and the duty cycle is set to $0.005$. Due to the substantial computational cost required by the $\Gaus_I$ model, we restrict this comparison to the $\Gaus$ and $\Gaus_C$ models. 

The parameter-estimation bias distributions obtained from 50 simulations are shown in Fig.~\ref{fig_histogram_bias_with_frequency}. For comparison, we also show the results from the simplified setup, namely those in Fig.~\ref{fig_histogram_bias_isotropic} for $\rho^2_{\seg}=0.1$. The results are broadly consistent with the behavior observed in Fig.~\ref{fig_histogram_bias_isotropic}. In particular, $\Gaus_C$ improves the estimation accuracy of $(\xi,\Omega_{\mathrm{ref}})$ relative to $\Gaus$. We find that the realistic setup yields broader bias distributions, especially for the $\Gaus$ model. This is likely due to the inclusion of additional noise-amplitude parameters, which introduce extra freedom in the inference and can increase the uncertainty in the other parameters.

We find that neither model introduces a significant bias in the estimation of the noise-amplitude parameters $A_{n_I}$. This is likely because $A_{n_I}$ is a global parameter, while the vast majority of frequency bins are noise dominated, allowing both models to constrain it accurately. In a realistic analysis, however, detector noise may be non-stationary and exhibit frequency-dependent deviations from a Gaussian spectrum, requiring a more sophisticated noise model and a joint treatment of signal and noise uncertainties. We leave such extensions to future work.

\end{appendix}

\bibliography{refs}

@article{LIGOScientific:2014pky,
    author = "Aasi, J. and others",
    collaboration = "LIGO Scientific",
    title = "{Advanced LIGO}",
    eprint = "1411.4547",
    archivePrefix = "arXiv",
    primaryClass = "gr-qc",
    doi = "10.1088/0264-9381/32/7/074001",
    journal = "Class. Quant. Grav.",
    volume = "32",
    pages = "074001",
    year = "2015"
}

@article{VIRGO:2014yos,
    author = "Acernese, F. and others",
    collaboration = "VIRGO",
    title = "{Advanced Virgo: a second-generation interferometric gravitational wave detector}",
    eprint = "1408.3978",
    archivePrefix = "arXiv",
    primaryClass = "gr-qc",
    doi = "10.1088/0264-9381/32/2/024001",
    journal = "Class. Quant. Grav.",
    volume = "32",
    number = "2",
    pages = "024001",
    year = "2015"
}

@article{KAGRA:2020tym,
    author = "Akutsu, T. and others",
    collaboration = "KAGRA",
    title = "{Overview of KAGRA: Detector design and construction history}",
    eprint = "2005.05574",
    archivePrefix = "arXiv",
    primaryClass = "physics.ins-det",
    doi = "10.1093/ptep/ptaa125",
    journal = "PTEP",
    volume = "2021",
    number = "5",
    pages = "05A101",
    year = "2021"
}

@article{KAGRA:2021duu,
    author = "Abbott, R. and others",
    collaboration = "KAGRA, VIRGO, LIGO Scientific",
    title = "{Population of Merging Compact Binaries Inferred Using Gravitational Waves through GWTC-3}",
    eprint = "2111.03634",
    archivePrefix = "arXiv",
    primaryClass = "astro-ph.HE",
    reportNumber = "LIGO-P2100239 ; Data release: https://zenodo.org/record/5655785, LIGO-P2100239",
    doi = "10.1103/PhysRevX.13.011048",
    journal = "Phys. Rev. X",
    volume = "13",
    number = "1",
    pages = "011048",
    year = "2023"
}

@article{LIGOScientific:2018mvr,
    author = "Abbott, B. P. and others",
    collaboration = "LIGO Scientific, Virgo",
    title = "{GWTC-1: A Gravitational-Wave Transient Catalog of Compact Binary Mergers Observed by LIGO and Virgo during the First and Second Observing Runs}",
    eprint = "1811.12907",
    archivePrefix = "arXiv",
    primaryClass = "astro-ph.HE",
    reportNumber = "LIGO-P1800307",
    doi = "10.1103/PhysRevX.9.031040",
    journal = "Phys. Rev. X",
    volume = "9",
    number = "3",
    pages = "031040",
    year = "2019"
}

@article{LIGOScientific:2016dsl,
    author = "Abbott, B. P. and others",
    collaboration = "LIGO Scientific, Virgo",
    title = "{Binary Black Hole Mergers in the first Advanced LIGO Observing Run}",
    eprint = "1606.04856",
    archivePrefix = "arXiv",
    primaryClass = "gr-qc",
    reportNumber = "LIGO-P1600088",
    doi = "10.1103/PhysRevX.6.041015",
    journal = "Phys. Rev. X",
    volume = "6",
    number = "4",
    pages = "041015",
    year = "2016",
    note = "[Erratum: Phys.Rev.X 8, 039903 (2018)]"
}

@article{LIGOScientific:2021usb,
    author = "Abbott, R. and others",
    collaboration = "LIGO Scientific, VIRGO",
    title = "{GWTC-2.1: Deep extended catalog of compact binary coalescences observed by LIGO and Virgo during the first half of the third observing run}",
    eprint = "2108.01045",
    archivePrefix = "arXiv",
    primaryClass = "gr-qc",
    reportNumber = "LIGO-P2100063",
    doi = "10.1103/PhysRevD.109.022001",
    journal = "Phys. Rev. D",
    volume = "109",
    number = "2",
    pages = "022001",
    year = "2024"
}

@article{LIGOScientific:2020kqk,
    author = "Abbott, R. and others",
    collaboration = "LIGO Scientific, Virgo",
    title = "{Population Properties of Compact Objects from the Second LIGO-Virgo Gravitational-Wave Transient Catalog}",
    eprint = "2010.14533",
    archivePrefix = "arXiv",
    primaryClass = "astro-ph.HE",
    reportNumber = "LIGO-P2000077",
    doi = "10.3847/2041-8213/abe949",
    journal = "Astrophys. J. Lett.",
    volume = "913",
    number = "1",
    pages = "L7",
    year = "2021"
}

@article{LIGOScientific:2016fpe,
    author = "Abbott, B. P. and others",
    collaboration = "LIGO Scientific, Virgo",
    title = "{GW150914: Implications for the stochastic gravitational wave background from binary black holes}",
    eprint = "1602.03847",
    archivePrefix = "arXiv",
    primaryClass = "gr-qc",
    reportNumber = "LIGO-P1500222",
    doi = "10.1103/PhysRevLett.116.131102",
    journal = "Phys. Rev. Lett.",
    volume = "116",
    number = "13",
    pages = "131102",
    year = "2016"
}

@article{LIGOScientific:2017zlf,
    author = "Abbott, Benjamin P. and others",
    collaboration = "LIGO Scientific, Virgo",
    title = "{GW170817: Implications for the Stochastic Gravitational-Wave Background from Compact Binary Coalescences}",
    eprint = "1710.05837",
    archivePrefix = "arXiv",
    primaryClass = "gr-qc",
    reportNumber = "LIGO-P1700272",
    doi = "10.1103/PhysRevLett.120.091101",
    journal = "Phys. Rev. Lett.",
    volume = "120",
    number = "9",
    pages = "091101",
    year = "2018"
}

@article{LIGOScientific:2025slb,
    author = "Abac, A. G. and others",
    collaboration = "LIGO Scientific, VIRGO, KAGRA",
    title = "{GWTC-4.0: Updating the Gravitational-Wave Transient Catalog with Observations from the First Part of the Fourth LIGO-Virgo-KAGRA Observing Run}",
    eprint = "2508.18082",
    archivePrefix = "arXiv",
    primaryClass = "gr-qc",
    reportNumber = "LIGO-P2400386",
    month = "8",
    year = "2025"
}

@article{Drasco:2002yd,
    author = "Drasco, Steve and Flanagan, Eanna E.",
    title = "{Detection methods for nonGaussian gravitational wave stochastic backgrounds}",
    eprint = "gr-qc/0210032",
    archivePrefix = "arXiv",
    doi = "10.1103/PhysRevD.67.082003",
    journal = "Phys. Rev. D",
    volume = "67",
    pages = "082003",
    year = "2003"
}

@article{LIGOScientific:2020ibl,
    author = "Abbott, R. and others",
    collaboration = "LIGO Scientific, Virgo",
    title = "{GWTC-2: Compact Binary Coalescences Observed by LIGO and Virgo During the First Half of the Third Observing Run}",
    eprint = "2010.14527",
    archivePrefix = "arXiv",
    primaryClass = "gr-qc",
    reportNumber = "P2000061",
    doi = "10.1103/PhysRevX.11.021053",
    journal = "Phys. Rev. X",
    volume = "11",
    pages = "021053",
    year = "2021"
}

@article{LIGOScientific:2025brd,
    author = "Abac, A. G. and others",
    collaboration = "LIGO Scientific, Virgo, KAGRA",
    title = "{GW241011 and GW241110: Exploring Binary Formation and Fundamental Physics with Asymmetric, High-spin Black Hole Coalescences}",
    eprint = "2510.26931",
    archivePrefix = "arXiv",
    primaryClass = "astro-ph.HE",
    reportNumber = "LIGO-P2500402",
    doi = "10.3847/2041-8213/ae0d54",
    journal = "Astrophys. J. Lett.",
    volume = "993",
    number = "1",
    pages = "L21",
    year = "2025"
}

@article{LIGOScientific:2025kry,
    author = "Abac, A. G. and others",
    collaboration = "LIGO Scientific, VIRGO, KAGRA",
    title = "{Cosmological and High Energy Physics implications from gravitational-wave background searches in LIGO-Virgo-KAGRA's O1-O4a runs}",
    eprint = "2510.26848",
    archivePrefix = "arXiv",
    primaryClass = "gr-qc",
    reportNumber = "LIGO-PP2500150",
    month = "10",
    year = "2025"
}

@article{Regimbau:2011rp,
    author = "Regimbau, Tania",
    title = "{The astrophysical gravitational wave stochastic background}",
    eprint = "1101.2762",
    archivePrefix = "arXiv",
    primaryClass = "astro-ph.CO",
    doi = "10.1088/1674-4527/11/4/001",
    journal = "Res. Astron. Astrophys.",
    volume = "11",
    pages = "369--390",
    year = "2011"
}

@article{Punturo:2010zz,
    author = "Punturo, M. and others",
    editor = "Ricci, Fulvio",
    title = "{The Einstein Telescope: A third-generation gravitational wave observatory}",
    doi = "10.1088/0264-9381/27/19/194002",
    journal = "Class. Quant. Grav.",
    volume = "27",
    pages = "194002",
    year = "2010"
}

@article{LISA:2024hlh,
    author = "Colpi, Monica and others",
    collaboration = "LISA",
    title = "{LISA Definition Study Report}",
    eprint = "2402.07571",
    archivePrefix = "arXiv",
    primaryClass = "astro-ph.CO",
    month = "2",
    year = "2024"
}

@article{Evans:2021gyd,
    author = "Evans, Matthew and others",
    title = "{A Horizon Study for Cosmic Explorer: Science, Observatories, and Community}",
    eprint = "2109.09882",
    archivePrefix = "arXiv",
    primaryClass = "astro-ph.IM",
    reportNumber = "CE-P2100003-v7, Cosmic Explorer technical report CE-P2100003-v6",
    month = "9",
    year = "2021"
}

@article{TianQin:2015yph,
    author = "Luo, Jun and others",
    collaboration = "TianQin",
    title = "{TianQin: a space-borne gravitational wave detector}",
    eprint = "1512.02076",
    archivePrefix = "arXiv",
    primaryClass = "astro-ph.IM",
    doi = "10.1088/0264-9381/33/3/035010",
    journal = "Class. Quant. Grav.",
    volume = "33",
    number = "3",
    pages = "035010",
    year = "2016"
}

@article{Ruan:2018tsw,
    author = "Ruan, Wen-Hong and Guo, Zong-Kuan and Cai, Rong-Gen and Zhang, Yuan-Zhong",
    title = "{Taiji program: Gravitational-wave sources}",
    eprint = "1807.09495",
    archivePrefix = "arXiv",
    primaryClass = "gr-qc",
    doi = "10.1142/S0217751X2050075X",
    journal = "Int. J. Mod. Phys. A",
    volume = "35",
    number = "17",
    pages = "2050075",
    year = "2020"
}

@article{Kawamura:2020pcg,
    author = "Kawamura, Seiji and others",
    title = "{Current status of space gravitational wave antenna DECIGO and B-DECIGO}",
    eprint = "2006.13545",
    archivePrefix = "arXiv",
    primaryClass = "gr-qc",
    doi = "10.1093/ptep/ptab019",
    journal = "PTEP",
    volume = "2021",
    number = "5",
    pages = "05A105",
    year = "2021"
}

@article{KAGRA:2021vkt,
    author = "Abbott, R. and others",
    collaboration = "KAGRA, VIRGO, LIGO Scientific",
    title = "{GWTC-3: Compact Binary Coalescences Observed by LIGO and Virgo during the Second Part of the Third Observing Run}",
    eprint = "2111.03606",
    archivePrefix = "arXiv",
    primaryClass = "gr-qc",
    reportNumber = "LIGO-P2000318",
    doi = "10.1103/PhysRevX.13.041039",
    journal = "Phys. Rev. X",
    volume = "13",
    number = "4",
    pages = "041039",
    year = "2023"
}

@article{Romano:2016dpx,
    author = "Romano, Joseph D. and Cornish, Neil J.",
    title = "{Detection methods for stochastic gravitational-wave backgrounds: a unified treatment}",
    eprint = "1608.06889",
    archivePrefix = "arXiv",
    primaryClass = "gr-qc",
    doi = "10.1007/s41114-017-0004-1",
    journal = "Living Rev. Rel.",
    volume = "20",
    number = "1",
    pages = "2",
    year = "2017"
}

@article{vanRemortel:2022fkb,
    author = "van Remortel, Nick and Janssens, Kamiel and Turbang, Kevin",
    title = "{Stochastic gravitational wave background: Methods and implications}",
    eprint = "2210.00761",
    archivePrefix = "arXiv",
    primaryClass = "gr-qc",
    doi = "10.1016/j.ppnp.2022.104003",
    journal = "Prog. Part. Nucl. Phys.",
    volume = "128",
    pages = "104003",
    year = "2023"
}

@article{Thrane:2013kb,
    author = "Thrane, Eric",
    title = "{Measuring the non-Gaussian stochastic gravitational-wave background: a method for realistic interferometer data}",
    eprint = "1301.0263",
    archivePrefix = "arXiv",
    primaryClass = "astro-ph.IM",
    doi = "10.1103/PhysRevD.87.043009",
    journal = "Phys. Rev. D",
    volume = "87",
    number = "4",
    pages = "043009",
    year = "2013"
}

@article{Lawrence:2023buo,
    author = "Lawrence, Jessica and Turbang, Kevin and Matas, Andrew and Renzini, Arianna I. and van Remortel, Nick and Romano, Joseph D.",
    title = "{A stochastic search for intermittent gravitational-wave backgrounds}",
    eprint = "2301.07675",
    archivePrefix = "arXiv",
    primaryClass = "gr-qc",
    doi = "10.1103/PhysRevD.107.103026",
    journal = "Phys. Rev. D",
    volume = "107",
    number = "10",
    pages = "103026",
    year = "2023"
}

@article{Matas:2020roi,
    author = "Matas, Andrew and Romano, Joseph D.",
    title = "{Frequentist versus Bayesian analyses: Cross-correlation as an approximate sufficient statistic for LIGO-Virgo stochastic background searches}",
    eprint = "2012.00907",
    archivePrefix = "arXiv",
    primaryClass = "gr-qc",
    doi = "10.1103/PhysRevD.103.062003",
    journal = "Phys. Rev. D",
    volume = "103",
    number = "6",
    pages = "062003",
    year = "2021"
}

@article{Allen:1997ad,
    author = "Allen, Bruce and Romano, Joseph D.",
    title = "{Detecting a stochastic background of gravitational radiation: Signal processing strategies and sensitivities}",
    eprint = "gr-qc/9710117",
    archivePrefix = "arXiv",
    reportNumber = "WISC-MILW-97-TH-14",
    doi = "10.1103/PhysRevD.59.102001",
    journal = "Phys. Rev. D",
    volume = "59",
    pages = "102001",
    year = "1999"
}

@article{Allen:1996gp,
    author = "Allen, Bruce and Ottewill, Adrian C.",
    title = "{Detection of anisotropies in the gravitational wave stochastic background}",
    eprint = "gr-qc/9607068",
    archivePrefix = "arXiv",
    reportNumber = "WISC-MILW-96-TH-15",
    doi = "10.1103/PhysRevD.56.545",
    journal = "Phys. Rev. D",
    volume = "56",
    pages = "545--563",
    year = "1997"
}

@article{Belczynski:2001uc,
    author = "Belczynski, Krzysztof and Kalogera, Vassiliki and Bulik, Tomasz",
    title = "{A Comprehensive study of binary compact objects as gravitational wave sources: Evolutionary channels, rates, and physical properties}",
    eprint = "astro-ph/0111452",
    archivePrefix = "arXiv",
    doi = "10.1086/340304",
    journal = "Astrophys. J.",
    volume = "572",
    pages = "407--431",
    year = "2001"
}

@article{Bethe:1998bn,
    author = "Bethe, Hans A. and Brown, G. E.",
    title = "{Evolution of binary compact objects which merge}",
    eprint = "astro-ph/9802084",
    archivePrefix = "arXiv",
    reportNumber = "SUNY-NTG-98-4",
    doi = "10.1086/306265",
    journal = "Astrophys. J.",
    volume = "506",
    pages = "780--789",
    year = "1998"
}

@book{Feller1971,
  author    = {Feller, William},
  title     = {An Introduction to Probability Theory and Its Applications},
  volume    = {II},
  edition   = {2},
  publisher = {Wiley},
  address   = {New York},
  year      = {1971},
  note      = {Chap.~XVI},
}

@article{Buscicchio:2022raf,
    author = "Buscicchio, Riccardo and Ain, Anirban and Ballelli, Matteo and Cella, Giancarlo and Patricelli, Barbara",
    title = "{Detecting non-Gaussian gravitational wave backgrounds: A unified framework}",
    eprint = "2209.01400",
    archivePrefix = "arXiv",
    primaryClass = "gr-qc",
    doi = "10.1103/PhysRevD.107.063027",
    journal = "Phys. Rev. D",
    volume = "107",
    number = "6",
    pages = "063027",
    year = "2023"
}

@article{Ballelli:2022bli,
    author = "Ballelli, Matteo and Buscicchio, Riccardo and Patricelli, Barbara and Ain, Anirban and Cella, Giancarlo",
    title = "{Improved detection statistics for non-Gaussian gravitational wave stochastic backgrounds}",
    eprint = "2212.10038",
    archivePrefix = "arXiv",
    primaryClass = "gr-qc",
    doi = "10.1103/PhysRevD.107.124044",
    journal = "Phys. Rev. D",
    volume = "107",
    number = "12",
    pages = "124044",
    year = "2023"
}

@article{LIGOScientific:2025bgj,
    author = "Abac, A. G. and others",
    collaboration = "LIGO Scientific, VIRGO, KAGRA",
    title = "{Upper Limits on the Isotropic Gravitational-Wave Background from the first part of LIGO, Virgo, and KAGRA's fourth Observing Run}",
    eprint = "2508.20721",
    archivePrefix = "arXiv",
    primaryClass = "gr-qc",
    reportNumber = "LIGO-P2500349",
    month = "8",
    year = "2025"
}

@article{Franciolini:2025leq,
    author = "Franciolini, Gabriele and Pieroni, Mauro and Ricciardone, Angelo and Romano, Joseph D.",
    title = "{Likelihoods for Stochastic Gravitational Wave Background Data Analysis}",
    eprint = "2505.24695",
    archivePrefix = "arXiv",
    primaryClass = "gr-qc",
    reportNumber = "CERN-TH-2025-094",
    month = "5",
    year = "2025"
}

@article{LIGOScientific:2025pvj,
    author = "Abac, A. G. and others",
    collaboration = "LIGO Scientific, VIRGO, KAGRA",
    title = "{GWTC-4.0: Population Properties of Merging Compact Binaries}",
    eprint = "2508.18083",
    archivePrefix = "arXiv",
    primaryClass = "astro-ph.HE",
    reportNumber = "LIGO-P2400004",
    month = "8",
    year = "2025"
}

@article{Lamb:2024gbh,
    author = "Lamb, William G. and Taylor, Stephen R.",
    title = "{Spectral Variance in a Stochastic Gravitational-wave Background from a Binary Population}",
    eprint = "2407.06270",
    archivePrefix = "arXiv",
    primaryClass = "gr-qc",
    doi = "10.3847/2041-8213/ad654a",
    journal = "Astrophys. J. Lett.",
    volume = "971",
    number = "1",
    pages = "L10",
    year = "2024"
}

@article{Coward:2006df,
    author = "Coward, David and Regimbau, Tania",
    title = "{Detection regimes of the cosmological gravitational wave background from astrophysical sources}",
    eprint = "astro-ph/0607043",
    archivePrefix = "arXiv",
    doi = "10.1016/j.newar.2006.07.001",
    journal = "New Astron. Rev.",
    volume = "50",
    pages = "461--467",
    year = "2006"
}

@article{Rosado:2011kv,
    author = "Rosado, Pablo A.",
    title = "{Gravitational wave background from binary systems}",
    eprint = "1106.5795",
    archivePrefix = "arXiv",
    primaryClass = "gr-qc",
    doi = "10.1103/PhysRevD.84.084004",
    journal = "Phys. Rev. D",
    volume = "84",
    pages = "084004",
    year = "2011"
}

@article{Ravi:2012bz,
    author = "Ravi, V. and Wyithe, J. S. B. and Hobbs, G. and Shannon, R. M. and Manchester, R. N. and Yardley, D. R. B. and Keith, M. J.",
    title = "{Does a 'stochastic' background of gravitational waves exist in the pulsar timing band?}",
    eprint = "1210.3854",
    archivePrefix = "arXiv",
    primaryClass = "astro-ph.CO",
    doi = "10.1088/0004-637X/761/2/84",
    journal = "Astrophys. J.",
    volume = "761",
    pages = "84",
    year = "2012"
}

@article{Regimbau:2009rk,
    author = "Regimbau, Tania and Hughes, Scott A.",
    title = "{Gravitational-wave confusion background from cosmological compact binaries: Implications for future terrestrial detectors}",
    eprint = "0901.2958",
    archivePrefix = "arXiv",
    primaryClass = "gr-qc",
    doi = "10.1103/PhysRevD.79.062002",
    journal = "Phys. Rev. D",
    volume = "79",
    pages = "062002",
    year = "2009"
}

@article{Braglia:2022icu,
    author = "Braglia, Matteo and Garcia-Bellido, Juan and Kuroyanagi, Sachiko",
    title = "{Tracking the origin of black holes with the stochastic gravitational wave background popcorn signal}",
    eprint = "2201.13414",
    archivePrefix = "arXiv",
    primaryClass = "astro-ph.CO",
    reportNumber = "IFT-UAM/CSIC-22-6",
    doi = "10.1093/mnras/stad082",
    journal = "Mon. Not. Roy. Astron. Soc.",
    volume = "519",
    number = "4",
    pages = "6008--6019",
    year = "2023"
}

@article{Bagui:2026hrf,
    author = "Bagui, Eleni and Clesse, S{\'e}bastien and De Lillo, Federico and Jenkins, Alexander C. and Sakellariadou, Mairi",
    title = "{Popcorn in the sky: Identifying primordial black holes in the gravitational-wave background}",
    eprint = "2601.07774",
    archivePrefix = "arXiv",
    primaryClass = "astro-ph.CO",
    month = "1",
    year = "2026"
}

@article{Yamamoto:2022kuh,
    author = "Yamamoto, Takahiro S. and Kuroyanagi, Sachiko and Liu, Guo-Chin",
    title = "{Deep learning for intermittent gravitational wave signals}",
    eprint = "2208.13156",
    archivePrefix = "arXiv",
    primaryClass = "gr-qc",
    doi = "10.1103/PhysRevD.107.044032",
    journal = "Phys. Rev. D",
    volume = "107",
    number = "4",
    pages = "044032",
    year = "2023"
}

@inproceedings{Utina:2021ipo,
    author = "Utina, Andrei and Marangio, Francesco and Morawski, Filip and Iess, Alberto and Regimbau, Tania and Fiameni, Giuseppe and Cuoco, Elena",
    title = "{Deep learning searches for gravitational wave stochastic backgrounds}",
    booktitle = "{International Conference on Content-Based Multimedia Indexing}",
    doi = "10.1109/CBMI50038.2021.9461904",
    month = "6",
    year = "2021"
}

@article{Smith:2017vfk,
    author = "Smith, Rory and Thrane, Eric",
    title = "{Optimal Search for an Astrophysical Gravitational-Wave Background}",
    eprint = "1712.00688",
    archivePrefix = "arXiv",
    primaryClass = "gr-qc",
    reportNumber = "LIGO-DOCUMENT-ID-LIGO-P1700407",
    doi = "10.1103/PhysRevX.8.021019",
    journal = "Phys. Rev. X",
    volume = "8",
    number = "2",
    pages = "021019",
    year = "2018"
}

@article{Smith:2020lkj,
    author = "Smith, Rory J. E. and Talbot, Colm and Hernandez Vivanco, Francisco and Thrane, Eric",
    title = "{Inferring the population properties of binary black holes from unresolved gravitational waves}",
    eprint = "2004.09700",
    archivePrefix = "arXiv",
    primaryClass = "astro-ph.HE",
    doi = "10.1093/mnras/staa1642",
    journal = "Mon. Not. Roy. Astron. Soc.",
    volume = "496",
    number = "3",
    pages = "3281--3290",
    year = "2020"
}

@article{Karnesis:2024pxh,
    author = "Karnesis, Nikolaos and Sasli, Argyro and Buscicchio, Riccardo and Stergioulas, Nikolaos",
    title = "{Characterization of non-Gaussian stochastic signals with heavier-tailed likelihoods}",
    eprint = "2410.14354",
    archivePrefix = "arXiv",
    primaryClass = "gr-qc",
    doi = "10.1103/PhysRevD.111.022005",
    journal = "Phys. Rev. D",
    volume = "111",
    number = "2",
    pages = "022005",
    year = "2025"
}

@article{Kou:2025bhk,
    author = "Kou, Xiao-Xiao and Saleem, Muhammed and Mandic, Vuk and Talbot, Colm and Thrane, Eric",
    title = "{Progress toward the detection of the gravitational-wave background from stellar-mass binary black holes: A mock data challenge}",
    eprint = "2506.14179",
    archivePrefix = "arXiv",
    primaryClass = "gr-qc",
    doi = "10.1103/h9w1-94m1",
    journal = "Phys. Rev. D",
    volume = "112",
    number = "8",
    pages = "084064",
    year = "2025"
}

@article{Biscoveanu:2020gds,
    author = "Biscoveanu, Sylvia and Talbot, Colm and Thrane, Eric and Smith, Rory",
    title = "{Measuring the primordial gravitational-wave background in the presence of astrophysical foregrounds}",
    eprint = "2009.04418",
    archivePrefix = "arXiv",
    primaryClass = "astro-ph.HE",
    reportNumber = "LIGO Document Number LIGO-P2000297",
    doi = "10.1103/PhysRevLett.125.241101",
    journal = "Phys. Rev. Lett.",
    volume = "125",
    pages = "241101",
    year = "2020"
}

@article{Dey:2023oui,
    author = "Dey, Ramit and Longo Micchi, Lu\'\i{}s Felipe and Mukherjee, Suvodip and Afshordi, Niayesh",
    title = "{Spectrogram correlated stacking: A novel time-frequency domain analysis of the stochastic gravitational wave background}",
    eprint = "2305.03090",
    archivePrefix = "arXiv",
    primaryClass = "gr-qc",
    doi = "10.1103/PhysRevD.109.023029",
    journal = "Phys. Rev. D",
    volume = "109",
    number = "2",
    pages = "023029",
    year = "2024"
}

@article{Sah:2023bgr,
    author = "Sah, Mohit Raj and Mukherjee, Suvodip",
    title = "{Non-Stationary Astrophysical Stochastic Gravitational-Wave Background: A New Probe to the High Redshift Population of Binary Black Holes}",
    eprint = "2307.06405",
    archivePrefix = "arXiv",
    primaryClass = "gr-qc",
    month = "7",
    year = "2023"
}

@article{Sah:2025agw,
    author = "Sah, Mohit Raj and Mukherjee, Suvodip",
    title = "{The First Upper Bound on the Non-Stationary Gravitational Wave Background and its Implication on the High Redshift Binary Black Hole Merger Rate}",
    eprint = "2511.03262",
    archivePrefix = "arXiv",
    primaryClass = "astro-ph.HE",
    month = "11",
    year = "2025"
}

@article{Seto:2008xr,
    author = "Seto, Naoki",
    title = "{Non-Gaussianity test for discriminating gravitational wave backgrounds around 0.1-1Hz}",
    eprint = "0807.1151",
    archivePrefix = "arXiv",
    primaryClass = "astro-ph",
    doi = "10.1086/591847",
    journal = "Astrophys. J. Lett.",
    volume = "683",
    pages = "L95--L98",
    year = "2008"
}

@article{Seto:2009ju,
    author = "Seto, Naoki",
    title = "{Non-Gaussianity analysis of GW background made by short-duration burst signals}",
    eprint = "0908.0228",
    archivePrefix = "arXiv",
    primaryClass = "gr-qc",
    doi = "10.1103/PhysRevD.80.043003",
    journal = "Phys. Rev. D",
    volume = "80",
    pages = "043003",
    year = "2009"
}

@article{Sasli:2023mxr,
    author = "Sasli, Argyro and Karnesis, Nikolaos and Stergioulas, Nikolaos",
    title = "{Heavy-tailed likelihoods for robustness against data outliers: Applications to the analysis of gravitational wave data}",
    eprint = "2305.04709",
    archivePrefix = "arXiv",
    primaryClass = "gr-qc",
    doi = "10.1103/PhysRevD.108.103005",
    journal = "Phys. Rev. D",
    volume = "108",
    number = "10",
    pages = "103005",
    year = "2023"
}

@article{Cornish:2015pda,
    author = "Cornish, Neil J. and Romano, Joseph D.",
    title = "{When is a gravitational-wave signal stochastic?}",
    eprint = "1505.08084",
    archivePrefix = "arXiv",
    primaryClass = "gr-qc",
    doi = "10.1103/PhysRevD.92.042001",
    journal = "Phys. Rev. D",
    volume = "92",
    number = "4",
    pages = "042001",
    year = "2015"
}

@article{Jaranowski:1998qm,
    author = "Jaranowski, Piotr and Krolak, Andrzej and Schutz, Bernard F.",
    title = "{Data analysis of gravitational - wave signals from spinning neutron stars. 1. The Signal and its detection}",
    eprint = "gr-qc/9804014",
    archivePrefix = "arXiv",
    doi = "10.1103/PhysRevD.58.063001",
    journal = "Phys. Rev. D",
    volume = "58",
    pages = "063001",
    year = "1998"
}

@article{Thrane:2009fp,
    author = "Thrane, Eric and Ballmer, Stefan and Romano, Joseph D. and Mitra, Sanjit and Talukder, Dipongkar and Bose, Sukanta and Mandic, Vuk",
    title = "{Probing the anisotropies of a stochastic gravitational-wave background using a network of ground-based laser interferometers}",
    eprint = "0910.0858",
    archivePrefix = "arXiv",
    primaryClass = "astro-ph.IM",
    doi = "10.1103/PhysRevD.80.122002",
    journal = "Phys. Rev. D",
    volume = "80",
    pages = "122002",
    year = "2009"
}

@article{vanDie:2024htf,
    author = "van Die, Frans and Rapoport, Ivan and Ginat, Yonadav Barry and Desjacques, Vincent",
    title = "{Detection prospects for the GW background of galactic (sub)solar mass primordial black holes}",
    eprint = "2410.04522",
    archivePrefix = "arXiv",
    primaryClass = "astro-ph.CO",
    doi = "10.1088/1475-7516/2025/05/036",
    journal = "JCAP",
    volume = "05",
    pages = "036",
    year = "2025"
}

@article{Afroz:2025urb,
    author = "Afroz, Samsuzzaman and Mukherjee, Suvodip",
    title = "{Gravitational Wave Burst from Bremsstrahlung in Milky Way Can Discover Sub-Solar Dark Matter in Near Future}",
    eprint = "2507.22126",
    archivePrefix = "arXiv",
    primaryClass = "astro-ph.CO",
    month = "7",
    year = "2025"
}

@article{Regimbau:2011bm,
    author = "Regimbau, Tania and Giampanis, Stefanos and Siemens, Xavier and Mandic, Vuk",
    title = "{The stochastic background from cosmic (super)strings: popcorn and (Gaussian) continuous regimes}",
    eprint = "1111.6638",
    archivePrefix = "arXiv",
    primaryClass = "astro-ph.CO",
    doi = "10.1103/PhysRevD.85.066001",
    journal = "Phys. Rev. D",
    volume = "85",
    pages = "066001",
    year = "2012"
}

@article{Renzini:2023qtj,
    author = "Renzini, Arianna I. and others",
    title = "{pygwb: A Python-based Library for Gravitational-wave Background Searches}",
    eprint = "2303.15696",
    archivePrefix = "arXiv",
    primaryClass = "gr-qc",
    doi = "10.3847/1538-4357/acd775",
    journal = "Astrophys. J.",
    volume = "952",
    number = "1",
    pages = "25",
    year = "2023"
}

@article{Buonanno:2004tp,
    author = "Buonanno, Alessandra and Sigl, Gunter and Raffelt, Georg G. and Janka, Hans-Thomas and Muller, Ewald",
    title = "{Stochastic gravitational wave background from cosmological supernovae}",
    eprint = "astro-ph/0412277",
    archivePrefix = "arXiv",
    doi = "10.1103/PhysRevD.72.084001",
    journal = "Phys. Rev. D",
    volume = "72",
    pages = "084001",
    year = "2005"
}

@article{Martellini:2014xia,
    author = "Martellini, Lionel and Regimbau, Tania",
    title = "{Semiparametric approach to the detection of non-Gaussian gravitational wave stochastic backgrounds}",
    eprint = "1405.5775",
    archivePrefix = "arXiv",
    primaryClass = "astro-ph.CO",
    doi = "10.1103/PhysRevD.89.124009",
    journal = "Phys. Rev. D",
    volume = "89",
    number = "12",
    pages = "124009",
    year = "2014"
}

@article{Martellini:2015mfr,
    author = "Martellini, Lionel and Regimbau, Tania",
    title = "{Efficiency of the cross-correlation statistic for gravitational wave stochastic background signals with non-Gaussian noise and heterogeneous detector sensitivities}",
    eprint = "1509.04802",
    archivePrefix = "arXiv",
    primaryClass = "astro-ph.CO",
    doi = "10.1103/PhysRevD.92.104025",
    journal = "Phys. Rev. D",
    volume = "92",
    number = "10",
    pages = "104025",
    year = "2015",
    note = "[Erratum: Phys.Rev.D 97, 049903 (2018)]"
}

@article{Buscicchio:2024wwm,
    author = "Buscicchio, Riccardo and Klein, Antoine and Korol, Valeriya and Di Renzo, Francesco and Moore, Christopher J. and Gerosa, Davide and Carzaniga, Alessandro",
    title = "{Test for LISA foreground Gaussianity and stationarity: galactic white-dwarf binaries}",
    eprint = "2410.08263",
    archivePrefix = "arXiv",
    primaryClass = "astro-ph.HE",
    doi = "10.1140/epjc/s10052-025-14616-w",
    journal = "Eur. Phys. J. C",
    volume = "85",
    number = "8",
    pages = "887",
    year = "2025"
}

@article{Ungarelli:2001xu,
    author = "Ungarelli, Carlo and Vecchio, Alberto",
    title = "{Studying the anisotropy of the gravitational wave stochastic background with LISA}",
    eprint = "astro-ph/0106538",
    archivePrefix = "arXiv",
    doi = "10.1103/PhysRevD.64.121501",
    journal = "Phys. Rev. D",
    volume = "64",
    pages = "121501",
    year = "2001"
}

@article{Weaver:2026qgr,
    author = "Weaver, Tristan S. and Radice, David and Jeong, Donghui and Liu, Victor",
    title = "{Gravitational Wave Modeling of White-Dwarf--Compact-Object Binaries and Observational Outlook}",
    eprint = "2607.24951",
    archivePrefix = "arXiv",
    primaryClass = "astro-ph.HE",
    month = "7",
    year = "2026"
}

@article{Adams:2013qma,
    author = "Adams, Matthew R. and Cornish, Neil J.",
    title = "{Detecting a Stochastic Gravitational Wave Background in the presence of a Galactic Foreground and Instrument Noise}",
    eprint = "1307.4116",
    archivePrefix = "arXiv",
    primaryClass = "gr-qc",
    doi = "10.1103/PhysRevD.89.022001",
    journal = "Phys. Rev. D",
    volume = "89",
    number = "2",
    pages = "022001",
    year = "2014"
}

@article{Chernoff:2009tp,
    author = "Chernoff, David F.",
    title = "{Clustering of Superstring Loops}",
    eprint = "0908.4077",
    archivePrefix = "arXiv",
    primaryClass = "astro-ph.CO",
    month = "8",
    year = "2009"
}
\end{document}